\documentclass[aps,10pt,showpacs,preprintnumbers,nofootinbib]{revtex4-1}
\usepackage{graphicx}
\usepackage{dcolumn}
\usepackage{bm,amsfonts,amsthm,amsmath,amssymb}
\usepackage[]{epsfig,graphics}
\usepackage{comment}
\usepackage[T1]{fontenc}
\usepackage{lipsum}
\usepackage[utf8]{inputenc}
\usepackage{ulem}
\usepackage{multirow}
\usepackage{color}
\usepackage{lastpage}
\usepackage{enumerate}
\usepackage{subfigure}
\usepackage{wasysym}
\usepackage{hyperref}
\usepackage{graphicx}
\usepackage{caption}

\hypersetup{
    bookmarks=true,         
    unicode=false,          
    pdftoolbar=true,        
    pdfmenubar=true,        
    pdffitwindow=false,     
    pdfstartview={FitH},    
    pdftitle={My title},    
    pdfauthor={Author},     
    pdfsubject={Subject},   
    pdfcreator={Creator},   
    pdfproducer={Producer}, 
    pdfkeywords={keyword1} {key2} {key3}, 
    pdfnewwindow=true,      
    colorlinks=false,       
    linkcolor=red,          
    citecolor=green,        
    filecolor=magenta,      
    urlcolor=cyan           
}

\newenvironment{itemize*}
  {\begin{itemize}
    \setlength{\itemsep}{0pt}
    \setlength{\parskip}{0pt}}
  {\end{itemize}}

\newenvironment{enumerate*}
  {\begin{enumerate}
    \setlength{\itemsep}{0pt}
    \setlength{\parskip}{0pt}}
  {\end{enumerate}}

\newenvironment{description*}
  {\begin{description}
    \setlength{\itemsep}{0pt}
    \setlength{\parskip}{0pt}}
  {\end{description}}

\def\ben{\begin{enumerate*}}
\def\een{\end{enumerate*}}
\def\bi{\begin{itemize*}}
\def\ei{\end{itemize*}}
\def\bd{\begin{description*}}
\def\ed{\end{description*}}
\def\be{\begin{equation}}
\def\ee{\end{equation}}
\def\bea{\begin{eqnarray}}
\def\eea{\end{eqnarray}}
\def\bfl{\begin{flushleft}}
\def\efl{\end{flushleft}}

\newcommand{\ttt}[1]{\mbox{\tiny #1}}
\newcommand{\mev}{\;\mbox{MeV}}
\newcommand{\gev}{\;\mbox{GeV}}

\newcommand{\gsim}{\lower.7ex\hbox{$\;\stackrel{\textstyle>}{\sim}\;$}}
\newcommand{\lsim}{\lower.7ex\hbox{$\;\stackrel{\textstyle<}{\sim}\;$}}
\newcommand{\sgm}{\sigma}
\newcommand{\beq}{\begin{equation}}
\newcommand{\eeq}{\end{equation}}
\newcommand{\fr}{\frac}
\newcommand{\qcd}{{\mbox{\tiny QCD}}}
\newcommand{\nn}{\nonumber}

\newcommand{\Sbar}{\overline{S}}

\begin{document}

\title{Cosmological Moduli and the Post-Inflationary Universe: \\ A Critical Review}
\author{Gordon Kane}
\email{gkane@umich.edu}
\affiliation{Michigan Center for Theoretical Physics, University of Michigan, Ann Arbor, MI 48109, USA}
\author{Kuver Sinha}
\email{kusinha@syr.edu}
\author{Scott Watson}
\email{gswatson@syr.edu} 
\affiliation{Department of Physics, Syracuse University, Syracuse, NY 13244, USA}

\date{\today}

\begin{abstract}
We critically review the role of cosmological moduli in determining the post-inflationary history of the universe.
Moduli are ubiquitous in string and M-theory constructions of beyond the Standard Model physics, where they parametrize the geometry of the compactification manifold.
For those with masses determined by supersymmetry breaking this leads to their eventual decay slightly before Big Bang Nucleosynthesis (without spoiling its predictions).   
This results in a matter dominated phase shortly after inflation ends, which can influence baryon and dark matter genesis, as well as observations of the Cosmic Microwave Background and the growth of large-scale structure. Given progress within fundamental theory, and guidance from dark matter and collider experiments, non-thermal histories have emerged as a robust and theoretically well-motivated alternative to a strictly thermal one. We review this approach to the early universe and discuss both the theoretical challenges and the observational implications.

\end{abstract}
\maketitle
\thispagestyle{empty}
\tableofcontents
\newpage

\section{Introduction}
The success of Big Bang cosmology in predicting the correct abundances of the light elements and the existence of the Cosmic Microwave Background (CMB)  rests upon the assumption that the universe began in a hot, dense state and as the universe expanded it cooled.  This {\it thermal history} for the early universe can then be used to try to understand the primordial origin of dark matter particles,  baryons, and the matter / anti-matter asymmetry.  This picture is simple, elegant, and robust, but it can not be complete.  Indeed, the primordial density perturbations -- indirectly detected through the temperature anisotropies of the CMB and required for the growth of the large-scale structure of the universe -- can not be causally generated in a strictly thermal universe.  Instead, inflation provides the most compelling model for their origin, and during this epoch the universe was not in thermal equilibrium. Moreover, following inflation the universe must {\it reheat} providing the hot big bang. This process can occur far from equilibrium and result in an effectively matter dominated epoch that could last up to the time of Big Bang Nucleosynthesis (BBN).  

Given the importance of realizing inflation and reheating, model building has had a rich and long history (for recent reviews see \cite{Baumann:2014nda, Baumann:2009ds}).  `Chaotic' inflation provides one of the simplest examples, where inflation is of the so-called large field type\footnote{Small field models of inflation typically require additional fine-tunings of initial conditions to enter the attractor regime (see e.g. \cite{Brandenberger:2003py} and references within.).  This tuning is in addition to that required for ensuring the flatness of the inflationary potential.} with the inflaton evolving over super-Planckian distances in the field space.  Some of these models can lead to detectable gravitational wave signals, but they are also sensitive to Planck scale physics and dangerous quantum gravity corrections. 
As a result, a large effort has been put forth to understand how to construct models in the controlled framework of string theory.  One lesson from these studies is that typically additional symmetries (such as Supersymmetry (SUSY)) and additional scalar fields (moduli) are generic and necessary ingredients for constructing realistic models. The former are necessary for protecting the inflaton potential from dangerous UV corrections\footnote{In string theory, large field models with approximate shift symmetries providing the flatness of the potential can be constructed, but additional symmetries -- such as SUSY -- are required to protect this potential from further corrections.}, whereas moduli are a ubiquitous prediction resulting from both the presence of these new symmetries, as well as the string construction itself.  If unstable these moduli can spoil inflation, or lead to undesirable late-time effects such as large entropy production after BBN or fifth force violations of General Relativity. Thus, although string theory provides a framework to construct realistic models of inflation, it appears to introduce a new nuisance -- the Cosmological Moduli Problem\footnote{The cosmological moduli problem made its first appearance in the form of the Polonyi problem in the earliest versions of spontaneously broken supergravity. This was the observation that hidden sector models of supersymmetry breaking with gravity mediation often contain scalar fields with weak scale masses and gravitational couplings. Just after inflation, these fields would behave like non-relativisitic matter and continue to dominate the energy density of the universe until after BBN.} \cite{Banks:1995dp,Banks:1995dt,Coughlan:1983ci,deCarlos:1993jw,Banks:1993en}.

Significant progress has been made on moduli stabilization both during and after inflation (for reviews see \cite{Grana:2005jc, Douglas:2006es}).  Many of the moduli have been shown to be stabilized with string scale masses by accounting for a combination of effects resulting from the presence of branes, fluxes, and/or strong hidden sector dynamics -- all of which are expected ingredients of the theory. However, in explicitly constructed examples it is also a common result that many of the moduli will remain parametrically lighter than the string scale.   The resulting cosmology differs from that of a strictly thermal post-inflationary history. These fields are displaced from their low-energy minima in the early universe and undergo coherent oscillations, mimicking a matter dominated epoch\footnote{This is strictly true only if the mass term in the potential gives the dominant contribution, otherwise the cosmological scaling of the energy density is determined by the dominant term in the potential~\cite{Turner:1983he}.} prior to BBN.
The fields typically decay late through gravitational strength couplings, and the universe reheats (again) via the production of relativistic Standard Model and other light (beyond the Standard Model) particles. Depending on the masses and couplings to other fields, such as dark matter and baryons, this can lead to a new expectation for the post-inflationary cosmology prior to BBN.

The rest of the paper is structured as follows.
In Section \ref{sec2}, we review the cosmological moduli problem and briefly comment on its appearance in effective supergravity descriptions of string / M-theory. In Section \ref{dm_section} we discuss the effect of cosmological moduli on the thermal history of dark matter. We look at the non-thermal WIMP miracle and its phenomenological status, as well as axions from string theory and their status as dark matter candidates within non-thermal histories. In Section \ref{baryosection}, we study the impact of cosmoligical moduli on theories of the baryon asymmetry of the universe, critically reviewing the Affleck-Dine mechanism and the baryon-dark matter cosmic coincidence. In Section \ref{cmblssDM}, we study cosmological moduli in the context of the CMB and Large Scale Structure, looking at implications for the matter power spectrum, effective number of neutrino species, and isocurvature constraints. We end with a section on the challenges and future prospects of these scenarios.

\section{String theory, Moduli, and a Non-thermal Universe \label{sec2}}
\subsection{Moduli Decay and Another Big Bang}
To see how a non-thermal history arises from moduli, we consider a single modulus\footnote{The case of non-thermal histories with multiple moduli was considered in \cite{Acharya:2008bk}.  Generically, string / M-theory theories have many moduli with various masses, and the lightest is typically the relevant one.} which enjoys a classical shift symmetry $\sigma \rightarrow \sigma + c$. 
At low energy we expect the symmetry to be broken, resulting in a mass for the field $m_\sigma$.  Moreover, if the presence of this new symmetry is tied to addressing the     
electroweak hierarchy problem (e.g. in SUSY or extra dimensional approaches) then we expect $m_\sigma \ll m_p$ with $m_p=2.4 \times 10^{18}$ GeV the reduced Planck mass. 
The non-thermal history arises from the observation that there is no {\it a-priori} reason why the modulus $\sigma$ should initially begin in its low-energy minimum. Both thermal and quantum fluctuations can displace the scalar during the early evolution leading to a matter dominated phase \cite{Dine:1995kz}. 

As an explicit example, consider string moduli in SUSY based approaches where on general grounds we expect that the shift symmetry of the modulus should be broken by both the finite energy density of inflation (which spontaneously breaks SUSY at the high energy) and quantum gravity effects \cite{Dine:1995kz}.  These considerations imply additional contributions to the effective potential in the form of a Hubble scale mass and a tower of non-renormalizable operators,
\be \label{potentialex}
\Delta V_1 = -c_1 H_{I}^2 \sigma^2 + \frac{c_{n}}{M^{n+2}} \sigma^{4+n} + \ldots,
\ee
where in the absence of special symmetries $c_1$ and $c_{n}$ are expected to be positive\footnote{One can argue rather generally for the positivity of $c_n$ based on causality, unitarity, and demanding that the model have a UV completion (although a few known counter-examples exist in gravitational systems).  We refer the reader to \cite{Adams:2006sv} for details. On the other hand, the choice of $c_1$ positive (leading to a tachyonic mass) is chosen to realize the general expectation that the high energy and low energy minima of the field would not generically be expected to coincide in a gravitational and/or time dependent background \cite{Dine:1995uk}. Ideally one would like to calculate this in an explicit, UV complete model.} order one constants, $H_I$ is the Hubble rate during inflation, and $M$ is the scale of new physics, e.g. quantum gravity. As this will be important later, we note that the dimension of the leading irrelevant operator that lifts the flat direction is model dependent as well as the scale of new physics\footnote{This is a generic expectation in string theories where new thresholds before the Planck scale are common place.  Examples include both the compactification (Kaluza-Klein) scale $M_{kk}$ and string scale $M_s$ where $M \ll M_{pl}$ is required for consistency of the effective theory \cite{Polchinski:1998rq}.}.
Because the contributions (\ref{potentialex}) are the dominant terms in the potential during high scale inflation ($H_I>m_\sigma$), this implies the minimum of the field at that time will be 
\be
\langle \sigma \rangle \sim M \left( \frac{H_I}{M} \right)^{\frac{2}{n+2}}\, .
\label{eqn:sig_vev_inf}
\ee
Whereas, later the low-energy SUSY breaking mass will give the dominant contributions to the potential, and this minimum is near $\langle \sigma \rangle \sim 0$.  This initial displacement from the low energy minimum eventually leads to coherent oscillations of the field and the formation of a scalar condensate. 
The amplitude of the oscillations is determined by the initial displacement $\sigma_\star \sim \langle \sigma \rangle$, where for the potential (\ref{potentialex}) we have $\sigma_\star \sim M \left( {H_I}/{M} \right)^{1 /(n+1)}$.  Hubble friction ceases and oscillations set in when the expansion rate satisfies $H \approx m_\sigma \ll H_I$.  Because the oscillations scale like matter, they dilute more slowly than the primordial radiation (produced during inflationary reheating).  Depending on the initial value $\sigma_\star$, the energy stored in the moduli may quickly come to dominate the energy density of the universe.  At the time oscillations begin $t_{\mbox{\tiny osc}} \approx H ^{-1} \approx m^{-1}_\sigma$ the initial abundance is given by
\be
\rho_\sigma(t_{\mbox{\tiny osc}}) = \frac{1}{2} m_\sigma^2  \sigma_\star^2,
\ee
and once the oscillations become coherent (which typically takes less than a Hubble time) they will scale as pressure-less matter \cite{Turner:1983he} with $\rho_\sigma \sim  m_\sigma^2  \sigma_\star^2 /a(t)^3$.  The universe remains matter dominated until the field decays.  Because the field is a modulus we expect it typically to be gravitationally coupled to other particles and so its decay rate is 
\be \label{decayrate} 
\Gamma_\sigma = \frac{c}{2\pi} \frac{m_\sigma^3}{\Lambda^2},
\ee
where we expect $\Lambda \sim M_{pl}$ and $c$ depends on the precise coupling in the fundamental Lagrangian, but typically takes values of at most ${\mathcal O}(100)$. 
Most of the field decays\footnote{In most of the literature the moduli decay is treated as instantaneous at $t_{\mbox{\tiny decay}}$.  However, this approximation can be misleading.  For example, in the case where the radiation energy density significantly drops below the moduli energy density ($\rho_\sigma \gg \rho_r$), the continuous (even though small) amounts of particle decay before reheating can lead to changes in the scale factor - temperature relation because the entropy is {\it not} conserved. We refer the reader to \cite{Giudice:2000ex} for further discussion.} at the time $t_{\rm decay} \sim H^{-1} \sim \Gamma_\sigma^{-1}$ and we expect it to decay democratically to Standard Model particles and their super-partners.  
Any heavy super-partners produced will typically decay rapidly into the lightest SUSY partner (LSP), which is stable and could provide a viable dark matter candidate (more on this in Section \ref{dm_section}). 

In addition to dark matter, light Standard Model particles that are produced will thermalize and `reheat' the universe for a second time (with inflationary reheating occurring early and at high temperature).  The corresponding reheat temperature is given by $T_r \sim g_*^{-1/4} \sqrt{\Gamma_\sigma \, M_{pl}} $ or
\be
 \label{Tr}
T_{\rm r} = c^{1/2} \left(\frac{10.75}{g_*}\right)^{1/4} \left( \frac{m_{\sigma}}{50\, {\rm TeV}}\right)^{3/2}\, T_{\rm BBN} \,, 
\ee
where $T_{\rm BBN} \simeq 3 ~ {\rm MeV}$ and $g_*$ is the number of relativistic degrees of freedom at $T_{\rm r}$,
and this temperature must be larger than around $3$ MeV to be in agreement with BBN light element abundances \cite{Kawasaki:1999na}. 
\begin{figure}[t]
\begin{center}
\includegraphics*[width=4.0in]{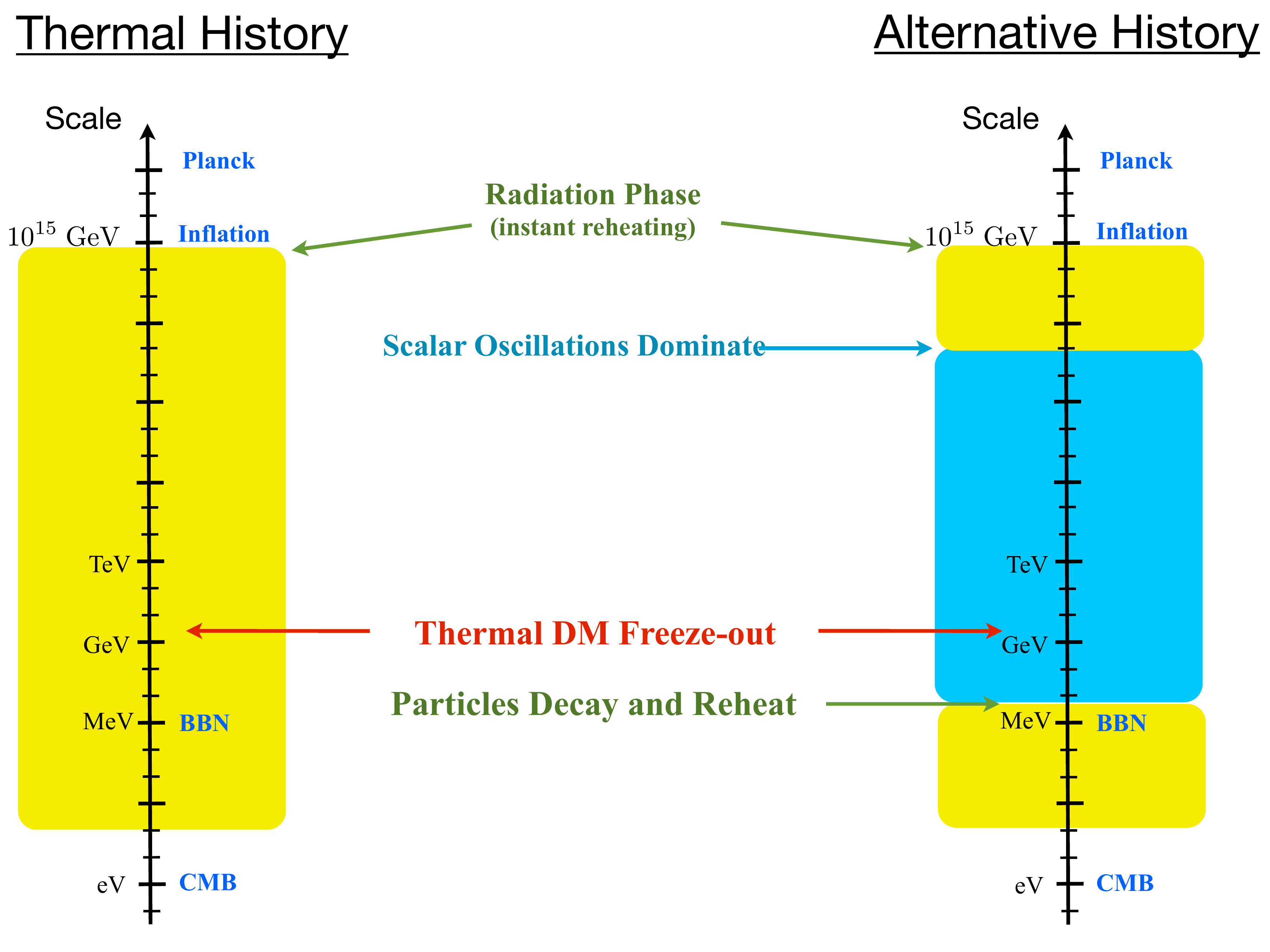} 
\hspace{0.2cm}
\end{center}
\caption{\label{fig1} The left timeline represents a thermal history for the early universe where dark matter WIMPs are populated in the thermal bath that emerges shortly after after inflationary reheating.  The right timeline represents a moduli dominated, non-thermal history, where the universe evolves as a matter dominated universe up until the decay time which must be before BBN.}
\end{figure}
 
A comparison between the standard thermal history and that of the non-thermal history appears in Figure \ref{fig1} for an interesting case where the modulus decays slightly before BBN and inflationary reheating occurs at a high temperature.  We now turn to the question of what moduli masses (and so reheat temperatures) are naturally expected from string / M-theory approaches.

\subsection{Fundamental Theory and Moduli Masses}
For an interesting departure from a thermal post-inflationary universe the 
mass of the lightest modulus should be in a window between $\mathcal{O}(10) - \mathcal{O}(1000)$ TeV.  For masses much further above the upper bound little departure from a thermal universe is expected, whereas the lower bound is required for consistency with BBN as seen from \eqref{Tr}. As a phenomenological approach
non-thermal cosmologies have been considered for their implications for dark matter (see \cite{Watson:2009hw} for a review).  However, lacking theoretical motivation these models are arguably somewhat exotic.

The key question is -- {\it should one generally expect a modulus in this mass range?} In string / M-theory approaches, inextricably connected with the question of moduli masses is the question of the gravitino mass. This is because moduli stabilization plays a central role in the nature of supersymmetry breaking. The potential of the moduli determines their F-terms dynamically, and along with the Kahler metric of the matter sector, determines the strength of gravity mediation. 
To extract the mass of the modulus, it is sufficient to write down its scalar potential in the supergravity limit (given in terms of its superpotential and Kahler potential, $W(\sigma)$ and $K(\sigma)$) 
\be \label{sp}
V=e^{K/m_p^2}  \left( \sum_\sigma \left| D_\sigma W  \right\vert^2 - \frac{3 |W|^2}{m_p^2} \right) \,\,.
\ee
One requires, at the extremum, that $\partial_{\sigma} V = 0$ and $\partial_{\sigma \sigma} V \, > \, 0$. Moreover, $V \sim 0$ is required to obtain an almost vanishing cosmological constant. We note that the gravitino mass is defined as $m^2_{3/2} = \langle e^{K} |W|^2 \rangle$ where we work in units of $m_p=1$ for simplicity.

This set of conditions is restrictive enough to give a relation between $m_{\sigma}$ and $m_{3/2}$, without going into the details of visible sector model-building. Since the modulus mass matrix is positive definite by assumption and both terms in \eqref{sp} are comparable to $m^2_{3/2}$ due to the condition $V \sim 0$, one obtains that the smallest eigenvalue of the mass matrix should be $\mathcal{O}(m_{3/2})$. For a single modulus, this trivially implies $\partial_{\sigma \sigma} V \, \sim \, m^2_{3/2}$. 

However, the modulus must be redefined to the physical basis with canonical kinetic term. An example Kahler potential is given by $K(\sigma) = -2 \ln \mathcal{V}$, where $\mathcal{V}$ is the volume form whose expression in terms of the moduli depends on the compactification manifold through the intersection numbers. In a basis with canonical kinetic terms, the lightest modulus will generally have a non-trivial profile along different field directions. However, for many settings that provide realistic phenomenology, the lightest modulus is also the largest, and one can approximate $\mathcal{V} \sim \sigma^{\alpha}$, where $\alpha \sim \mathcal{O}(1)$. This is the case in single modulus examples like KKLT \cite{Kachru:2003aw}, as well as the cases of Large Volume Compactifications in Type IIB \cite{Conlon:2013isa} and $G_2$ compactifications of M-theory \cite{Acharya:2008zi}. In that case, the canonical normalization yields $m_{\sigma} \sim \langle \sigma \rangle \partial_{\sigma \sigma} V$, where by abuse of notation $\sigma$ is now the physical field. One thus obtains
\be \label{modgravcorr}
m_{\sigma} \, \sim \, \langle \sigma \rangle \, m_{3/2} \,\,.
\ee

Usually, one finds $\langle \sigma \rangle \sim \mathcal{O}(10)-\mathcal{O}(100)$ in the supergravity regime. For example, if the modulus is stabilized supersymmetrically\footnote{The relation between moduli and gravitino mass in this case was first noted in \cite{LoaizaBrito:2005fa}.} so that $DW \sim 0$, then $\langle \sigma \rangle \sim \log \left( \frac{m_p}{m_{3/2}}  \right) \sim 40$. Then, under quite general circumstances one expects
\be \label{modgravcorr2}
m_{\sigma} \sim \mathcal{O}(1-100) \, \times \, m_{3/2} \,\, .
\ee
For $m_{3/2} \sim 10-100$ TeV, as one would expect in models that at least partially address the electroweak hierarchy problem, we see that $m_{\sigma}$ falls within the mass range where a non-thermal history results.

We would like to emphasize the interesting result that {\bf the lower bound on the mass required to address the  cosmological moduli problem provides further motivation for a little hierarchy between the electroweak scale and the scale of new scalar physics}.  Indeed, the  
discovery of a $125$ GeV Higgs at the Large Hadron Collider (LHC) -- and nothing else -- appears to have left 
the relevance of supersymmetry (SUSY) in question, particularly as a mechanism for stabilizing the hierarchy between the Electroweak and Planck scales.   Standard Model
Light superpartners (`Natural') SUSY still remains possible, but with some cost of increasing the level of complexity of models (see e.g.  \cite{Randall:2012dm}). Within the Minimal Supersymmetric Standard Model (MSSM), there are a number of alternative ways to reconcile SUSY with the data, including models of Split-SUSY \cite{Wells:2003tf,ArkaniHamed:2004fb,Arvanitaki:2012ps}, or simply by accepting that some fine-tuning may just be an accident of nature\footnote{It is noteworthy that even defining the level of fine-tuning can be an issue, see e.g. \cite{Baer:2013gva}.}. 
 Regardless of one's viewpoint, it seems that if low-energy SUSY will prevail it will require the existence of a new scale at around $10-100$ TeV.  
 
It is then remarkable that in these string / M-theory based approaches with moduli that the scale required to avoid the cosmological moduli problem is of the correct order of magnitude for realistic particle phenomenology\footnote{We note that already in 2006 this mass range for the lightest scalars and gravitino was noted in \cite{Acharya:2006ia} long before the Higgs discovery and LHC bounds favoring heavy scalar superpartners.} with a $125$ GeV Higgs.

Another interesting result of this string-based framework is that {\bf the reheat temperature is not a free parameter, but  a consequence of the hierarchy between the electroweak and Planck scale} (determined by $\Lambda_{susy}^2=m_{3/2} m_p$), which also helps determine the SUSY breaking masses of other sparticles in the theory. In both supergravity and string motivated approaches, the key lesson is that the reheat temperature is intimately connected to other aspects of the theory and not a free and tunable parameter.  Given that gravitationally coupled scalars or moduli are generic in  high energy UV completions of the Standard Model and in no sense exotic, we see that non-thermal histories are a feasible and robust possibility, and perhaps an inevitable one.  Given this motivation from theory, we now turn to the observational consequences of non-thermal histories.

\section{Dark Matter Genesis \label{dm_section}}

The origin of dark matter as a thermal relic is a simple and predictive scenario for its creation, which is also independent of the prior thermal history of the Universe. However, as we have argued so far, one is led on compelling theoretical grounds to a re-evaluation of the history of the Universe prior to BBN. This necessitates looking afresh at the consequences of non-thermal histories on dark matter phenomenology, a topic to which we turn in this section. In our choice of dark matter candidates, we will discuss both Weakly Interacting Massive Particles (WIMPs) and axions.

\subsection{WIMPs}

If the primordial universe is thermalized, massive long-lived particles may freeze-out with a final abundance determined primarily by their mass and annihilation cross-section \cite{Bertone:2004pz}. This is the basis of thermal WIMPs, which assumes a weak-scale cross-section, $\sigma$ and $\langle \sigma v \rangle_{th} \simeq10^{-26}$ cm$^3/$s, where  $v$ is the typical velocity.  Alternatively,   {\it non-thermal} dark matter is produced via the decay of heavier particles  into a long-lived final state and does not  require thermal equilibrium~\cite{Moroi:1999zb,Acharya:2008bk,Chung:1998rq,Fornengo:2002db,Pallis:2004yy,Gelmini:2006pw,Gelmini:2006pq} (for reviews, see \cite{Watson:2009hw, Baer:2014eja, Sinha:2013sxa}, and for recent related work \cite{Allahverdi:2013tca,Hooper:2011aj,Kelso:2013paa}).   Dark matter models are constrained by both direct detection experiments and searches for  astrophysical signals generated by their annihilation products. Non-thermal dark matter can have a higher annihilation cross-section than thermal dark matter so astrophysical signals are potentially stronger for these scenarios, particularly  in indirect experiments such as FERMI and AMS-2 \cite{Gelmini:2008sh,Grajek:2008pg,Grajek:2008jb,Dutta:2009uf,Kane:2009if,Sandick:2011rp}.  We now review the thermal and non-thermal WIMP paradigms.

\subsubsection{Thermal WIMPs}

The standard picture of the cosmic history assumes that the universe is radiation dominated prior to Big Bang Nucleosynthesis (BBN) and that dark matter is created from the thermal bath created at the end of inflation. The evolution of dark matter is described by the Boltzmann equation
\be \label{wimp1}
\dot{n}_X = -3 H n_X - \langle \sigma v \rangle \left[ n_X^2 - n_{eq}^2 \right],
\ee
where $\langle \sigma v \rangle$ is the thermally averaged cross-section, $n_X$ is the number
density, and $n_{eq}$ is the number density of the species in chemical equilibrium, i.e.
$XX \leftrightarrow \gamma \gamma$, where $\gamma$ is a relativistic particle. In the early universe with $T > m_X$, the dark matter is relativistic and in thermal equilibrium with the plasma, and its density is determined by the full solution to \eqref{wimp1}. When $T < m_X$, it becomes Boltzmann suppressed, until finally the expansion rate becomes larger than the annihilation rate, leading to thermal `freeze-out'. The dark matter density at this time is sensitive only to its mass and annihilation cross-section. 

Thus, the thermal paradigm is simple, predictive, and more or less independent of the underlying microscopic physics. Moreover, there is the suggestive fact that the observed relic density is satisfied for $\langle \sigma_xv \rangle \approx 10^{-26} $cm$^{3} \cdot $s$^{-1}$ (or $\sigma_x\approx 1$ picobarn), which corresponds to weak scale interactions. The weak scale is precisely where new physics is expected to appear, on very different physical grounds. Plausibly, then, dark matter is a Weakly Interacting Massive Particle (WIMP) that is there as a part of new physics at the weak scale, motivated by issues like the gauge hierarchy problem. The thermal WIMP miracle is precisely this combination of simplicity and serendipity. Since the weak scale is also the energy scale that we are currently able to probe experimentally, the thermal WIMP is by far the most actively studied dark matter candidate.

\subsubsection{The Non-thermal WIMP Miracle}

Despite its appeal, we saw in Section \ref{sec2} that there are compelling reasons to expect that the history of the universe prior to BBN was not as simple as the thermal WIMP paradigm requires. It is worthwhile to revisit the assumptions underlying the arguments above. They are: $(i)$ WIMPs reached chemical equilibrium in the early universe; $(ii)$ the universe was radiation dominated at the time of freeze-out; $(iii)$ following freeze-out, there was no dilution of the relic density due to entropy production from any sources, and conversely, there was no further production of dark matter from late-time sources.

The above assumptions look difficult to realize from a top-down perspective (and are in increasing tension with data from the bottom-up perspective, as we will discuss in Section~\ref{bottomup}).  Indeed, as we discussed in Section \ref{sec2} assumptions such as WIMPs achieving thermal equilibrium, the absence of entropy production following thermal freeze-out, and strictly a radiation dominated universe prior to BBN seem to be in conflict with top-down approaches to model building. 
In particular, the presence of moduli imply violations of all these conditions.

This by itself does not mean, however, that the thermal WIMP paradigm itself is disfavored, although a purely thermal history of the universe prior to BBN is. Moduli may decay \textit{early enough} (above the freeze-out temperature of WIMPs) to render a prior non-thermal history irrelevant as far as low-energy observables are concerned. The entropy dilution from such very early decay is likely to erase the prior dark matter density and produce it afresh (by reheating the modulus into the visible sector generally). The dark matter thus produced will then have time to thermalize and give the usual thermal WIMP story. 

A non-thermal WIMP is obtained if a modulus decays \textit{after} the freeze-out temperature but \textit{before} the onset of BBN. This translates into a window in terms of the decay width, and hence in terms of parameters of the fundamental theory (such as the modulus mass and overall decay constant). Leading up to \eqref{modgravcorr2} we have argued that it is natural for moduli to fall within this mass window, and in fact the explicitly studied compactifications in the literature have such moduli.

\subsubsection{Reheating Again and Dark Matter Production}

Moduli decay dilutes any previous population of dark matter by a factor $\Omega_{cdm} \rightarrow \Omega_{cdm} (T_r/T_f)^3$, which is typically large enough to render it irrelevant. This scaling can be understood from the fact that the temperature goes\footnote{If there is significant decay of the modulus during the moduli dominated epoch  entropy is not conserved and this alters the temperature / scale factor relation \cite{Giudice:2000ex}. In this review we will work in the instantaneous decay approximation and refer the reader to \cite{Fan:2014zua} for a more rigorous treatment. We note that for our considerations here we do not expect any significant qualitative differences.} as $T \sim 1/a(t)$, while volumes scale as $\sim a(t)^3$ (where $a(t)$ is the scale factor). 
\\

\noindent
The moduli decay is expected to produce dark matter and as a result there are two options:

$(1)$ If the WIMP number density exceeds the critical value $n^c_x =\left. \frac{H}{ \langle \sigma_xv \rangle} \right\vert_{T=T_r}$, then the WIMPs will quickly annihilate down to this value, which acts as an attractor \cite{Acharya:2008bk}. The fixed point value is evaluated at the reheating temperature instead of the freeze-out temperature. This results in a parametric enhancement of the relic density 
\be \label{TandNTrelation}
\Omega^{NT}_{dm} \, = \, \Omega^{T}_{dm} \left( \frac{T_f}{T_r} \right) \,\,.
\ee
Since the relic density scales inversely with the annihilation cross section, this implies that WIMP candidates with annihilation cross sections larger than the canonical value by a factor $(T_f/T_r)$ can give the correct relic density in a non-thermal history with the above compensating factor. Examples of such WIMPs are the Winos and Higgsinos of the minimal supersymmetric extension of the Standard Model (MSSM). For future reference, we will call this option the 'annihilation' scenario.

In more detail, the standard expression for the thermal relic density given by
\be  \label{thermal1}
\Omega_{dm}^{T}h^2 \, = \, \frac{45}{2 \pi \sqrt{10}} \left(\frac{s_0}{\rho_c m_p}\right) \left(
\frac{m_X }{g^{1/2}_* \langle \sigma v \rangle T_f} \right)
\ee
Using (\ref{TandNTrelation}), (\ref{thermal1}) and (\ref{nt1}) one can estimate the relic density in non-thermal dark matter as
\bea  \label{nt1}
\Omega_{dm}^{NT}h^2 &\simeq& 0.76 \, \left(\frac{s_0}{\rho_c m_p}\right) \left(
\frac{m_X }{g^{1/2}_* \langle \sigma v \rangle m_p T_r} \right) \, , \nonumber \\
&\simeq& 0.10 \, \left( \frac{m_X}{100 \; \mbox{GeV}}\right) \left(\frac{10.75}{g_\ast} \right)^{1/4} \left( \frac{3 \times 10^{-24} \; \mbox{cm}^3/\mbox{s}}{\langle \sigma v \rangle} \right) \left( \frac{\mbox{100 \,\mbox{TeV}}}{m_\sigma}\right)^{3/2},
\eea
where in the last line we used the entropy density today $s_0 = 2.78 \times 10^8 \rho_c/h^2$ with $h$ the Hubble parameter in units of $100$ km/s/Mpc and we chose some fiducial values with $g_{*} = 10.75$ the number of relativistic degrees of freedom at the time of reheating.
We note that unlike the thermal result \eqref{nt1} depends on both the properties of the dark matter (mass and annihilation rate) and on the mass of the decaying modulus ($m_\sigma$).
 As discussed in Section \ref{sec2} the scalar mass (and so reheat temperature) is {\it not} a free parameter, but related to the gravitino mass.
The mass of the scalar (and so the relic density of dark matter) is controlled by the need for SUSY to generate a hierarchy between the electroweak and Planck scale.
We have chosen a fiducial value for the annihilation rate that yields roughly the right amount of dark matter  for the  hierarchy set by the choice of low-scale SUSY breaking.  The  cross-section is three orders of magnitude higher than expected with a thermal history with important experimental consequences, as discussed next in Section \ref{bottomup}.

$(2)$ The second option we will call the `branching scenario'.  In this case, the WIMP number density produced from moduli decay is below the fixed point value. 
This implies that there is no further WIMP annihilations and the final number density is that produced from the decay, i.e., $n_x \sim B_x n_\sigma $, where $B_x$ is the branching ratio for scalar decay to WIMPs and $n_\sigma $ is the number density of the scalar condensate. With an appropriate choice of $n_x$ (that very much depends on the microphysics), this option can accommodate WIMPs whose annihilation cross section is higher and (crucially) also lower than the canonical value. Note that the latter case (a pure Bino of the MSSM) cannot be accommodated in any thermal setting without substantial fine-tuning of the spectrum. 

\subsubsection{Phenomenological Aspects and Future Probes} \label{bottomup}

We have seen that in theories with moduli and TeV scale SUSY breaking that a non-thermal history is attractive. When considering the associated phenomenology, it is non-trivial to avoid this conclusion without introducing fine-tuning. For example, one could raise both the gravitino and modulus mass, re-introducing the gauge hierarchy fine-tuning, or one could decouple the modulus-gravitino correlation in \eqref{modgravcorr} making the modulus much heavier as in racetrack models \cite{Kallosh:2004yh, BlancoPillado:2004ns, Linde:2007jn, Allahverdi:2009rm}. This introduces large fine-tuning that is not associated with the cosmological constant. 

From a bottom-up perspective, one of the most striking aspects of the thermal WIMP paradigm is how difficult it is to realize even with the large number of parameters within the MSSM. Taking Neutralinos as the WIMP candidates, the Wino and Higgsino -- which annihilate very efficiently -- must be very heavy ($2.5$ TeV and $1$ TeV, respectively) to give the correct relic density through thermal freeze-out \cite{ArkaniHamed:2006mb}.  Whereas a thermal Bino typically gives a relic density that is too high, unless its annihilation cross section is boosted through major co-annihilation effects \cite{Griest:1990kh} or mixing with the Higgsino \cite{ArkaniHamed:2006mb}. Both of these options are fine-tuned and the latter is highly constrained by direct detection  \cite{Hooper:2013qjx, Anandakrishnan:2014fia}. 
However, within non-thermal histories Wino or Higgsino dark matter with masses of $\mathcal{O}(100)$ GeV can be accommodated within the annihilation scenario discussed above.  Moreover, Bino dark matter with a non-thermal history following the 'branching scenario' can satisfy the relic density constraint, without relying on any other low-energy ingredient. 
Thus, an experimental confirmation of a pure gauge eigenstate of the MSSM as the dark matter of the universe would seem to suggest that the universe underwent a non-thermal history, since otherwise one would need to accept a large amount of fine-tuning. 
Direct detection limits are currently unable to probe such pure eigenstates, whose scattering cross-section off of Standard Model nuclei is typically $\sim \mathcal{O}(10^{-47})$ cm$^2$, although future facilities like XENON1T and LZ will reach the required sensitivity up to WIMP masses of $\sim 1$ TeV \cite{Cushman:2013zza}. 

Indirect detection provides an enormous amount of information by putting upper bounds on the dark matter annihilation cross section. A joint analysis of seven Milky Way dwarf galaxies using a frequentist Neyman construction and Pass 7 data from the Fermi Gamma-ray Space Telescope undertaken in \cite{Geringer-Sameth:2014qqa, GeringerSameth:2011iw} excludes generic thermal WIMP candidates with mass less than $\sim \, 40$ GeV annihilating to $b \bar{b}$ final states\footnote{The exclusion bounds are similar for other final states.}. Current limits using dwarf spheroidals data and the PASS-8 software from Fermi-LAT collaboration rule out the canonical WIMP for masses below $\sim 100$~GeV. WIMPs in this mass range thus over-satisfy the relic density in a purely thermal history. Since multi-component dark matter is not an option in this case, the only way WIMPs lighter than $\sim 100$ GeV can constitute the dark matter of the universe is if they followed a non-thermal 'branching scenario' cosmological history \cite{Allahverdi:2012wb}.  As this exclusion reach becomes larger, it forces more and more potential candidates squarely into the non-thermal framework. 

On the other hand, candidates like Winos and Higgsinos with large annihilation cross section are now severely bounded by a combination of continuum and line searches from FERMI data \cite{Fan:2013faa, Cohen:2013ama}. Continuum photons arise from the tree level annihilation process $\chi^0 \chi^0 \rightarrow W^{+} W^{-}$, where $\chi^0$ denotes the WIMP. Dwarf galaxy data rules out Wino dark matter masses up to around $400$ GeV, while galactic center data rules it out up to around $700$ GeV for either NFW or Einasto profiles \cite{Hryczuk:2014hpa}. These bounds become stronger (weaker) if one considers steeper (core) profiles.  Recent discoveries of optical signatures of black holes in dwarf galaxies indicates that low-mass galaxies can host intermediate massive black holes. Accounting for this without spoiling the measured velocity dispersion of stars, the constraints on the dark matter annihilation cross section may become orders of magnitude stronger \cite{Gonzalez-Morales:2014eaa}, ruling out thermal Wino dark matter masses up to 1 TeV. Similar exclusion bounds hold for any general scalar dark matter candidate that is a triplet under $SU(2)$ \cite{Queiroz:2014pra}.

CMB measurements are another source of constraint for non-thermal histories. We reserve a full discussion of this for Section \ref{cmblssDM}, here we  briefly mention how CMB observations can lead to constraints on the self-annihilation rate of non-thermal dark matter.  These constraints arise because 
energy injection from increased dark matter annihilation can alter the recombination history\footnote{This is the period during which the universe cools to a level where CMB photons can free-stream and no longer interact with electrons on average.} leading to changes in the temperature and polarization power spectra of the CMB \cite{Slatyer:2009yq}. We have seen that non-thermal dark matter can have a larger annihilation rate than thermal WIMPs, and so this would lead to stronger CMB constraints from the additional energy injection. 
The 2015 results from the PLANCK mission \cite{Planck:2015xua} imply a constraint on non-thermal WIMPs with masses 
$\sim 100$ GeV of $f_{eff} \langle \sigma v \rangle \lesssim 4 \times 10^{-25}$ cm$^3$ s$^{-1}$.  The constant $f_{eff}$ is a model dependent parameter that depends on the mass of the dark matter and captures the efficiency of the annihilations -- it typically takes values $f_{eff} \simeq 0.1$ -- $1$.  Although this parameter does introduce 
model dependence and some uncertainty into the constraints, with the 2015 Planck release it is already possible to place reasonable constraints on the annihilation rates of low-mass thermal WIMPs. Before reaching the cosmic variance limit, this type of constraint will be capable of ruling out non-thermal WIMPs completely when they have 
electroweak scale masses and comprise all of the cosmological dark matter.

Current collider bounds on Winos and Higgsinos are rather weak, given their low production cross-section at a hadron collider. 
Approaches to study the dark matter parameter space include effective operators \cite{Goodman:2010ku} and simplified models \cite{Abdallah:2014hon}, with a mono-$X$ observational signature, where $X$ may denote monojet \cite{Beltran:2010ww}, mono-photon \cite{Gershtein:2008bf}, mono-$Z$ \cite{Petriello:2008pu, Carpenter:2012rg}, mono-$W$ \cite{Bai:2012xg}, mono-Higgs, mono-$b$ \cite{Lin:2013sca}, or mono-top \cite{Lin:2013sca}. A particularly promising approach to the study of dark matter is the use of vector boson fusion processes \cite{Delannoy:2013ata, Cirelli:2014dsa, Baer:2014kya, Gori:2014oua, Dutta:2012xe, Dutta:2013gga, Giudice:2010wb}. The current exclusion prospects for pure Winos and Higgsinos are the following \cite{Berlin:2015aba}: taking a background systematic uncertainty of $1\% (5\%)$, with $3000$ fb$^{-1}$ of data, the 14 TeV LHC is sensitive to Winos of 240 GeV (125 GeV) and Higgsinos of 125 GeV (55
GeV). A future 100 TeV collider with 3000 fb$^{-1}$ of data would exclude Winos of mass 1.1 TeV (750 GeV) and Higgsinos of mass 530 GeV (180 GeV) taking $1\% (5\%)$ background uncertainty.

The above limits are in the case when everything except the lightest neutralino dark matter is decoupled from the low-energy spectrum. In specific models of string/M-theory compactifications, other light fields may be present. For example, in cases where other species such as a neutralino NLSP or gluino are accessible, the dark matter exclusion limits would depend on the final states and the decay modes of the NLSP or the colored state to the dark matter candidate. This is clearly more model-dependent. We briefly describe the low-energy spectrum and collider prospects of three of the most well-studied compactification schemes with non-thermal histories: mirage mediation models based on KKLT, Large Volume Scenarios, and the G2-MSSM.

In the case of mirage mediation models based on KKLT compactification, the dark matter candidate may be the Bino, Wino, or Higgsino, depending on the ratio (commonly denoted in the literature by $\alpha$) of the gravitino mass and the universal gaugino masses at the GUT scale. The mass spacing between the gauginos is determined by the same quantity, which also gives the relative strengths of the moduli and anomaly-mediated supersymmetric contributions. The rest of the spectrum depends on the modular weights of the visible sector. The collider and dark matter phenomenology of these models has been studied most recently in the post-Higgs era in \cite{Kaufman:2013oaa} and previously\footnote{These models have also been studied in the context of $SO(10)$ Yukawa-unified models \cite{Anandakrishnan:2013tqa, Anandakrishnan:2013cwa}.} in \cite{Dutta:2011kp, Baer:2007eh}. The canonical case of $\alpha = 1$ results in a spectrum with decoupled Bino dark matter, which is largely inaccessible in colliders and may be probed in future direct detection experiments. However, depending on the model parameters (modular weights) a gluino with mass $\sim 2300$ GeV may also be obtained in the spectrum in which case a discovery may be made with 300 fb$^{-1}$ of data at LHC-14. Changing the uplifting sector leads to other values of $\alpha$ which result in either decoupled Higgsinos or Winos with collider prospects described above.

In the case of Large Volume Scenarios, a wide range of phenomenological possibilities are present depending on the local construction of the visible sector (specifically the matter Kahler metric) and the way to achieve a de Sitter vacuum, as found in \cite{Aparicio:2014wxa}. In particular, two classes of spectra were obtained by these authors. The first is a split-SUSY spectrum with a decoupled scalar sector, whose collider prospects have been discussed earlier. The second is a CMSSM/mSUGRA scenario, which has been studied in detail in \cite{Aparicio:2015sda}. There it was shown that the confluence of collider bounds (LEP, LHC), CMB constraints, direct detection (LUX, XENON100, CDMS, IceCube) and indirect detection (Fermi) bounds result in an allowed parameter space that is characterized by  decoupled Higgsino-like dark matter with a mass around 300 GeV.

In the case of the G2-MSSM, the collider prospects have been studied most recently by the authors of \cite{Ellis:2014kla}. For a benchmark spectrum with a gravitino mass of $m_{3/2} = 35$ TeV, the gluino mass is expected to be about 1.5 TeV, while the Wino (Bino)-like gaugino mass is about 614 (450) GeV. Three and only three production channels should be discoverable at LHC-14: $pp \rightarrow \tilde{g} \tilde{g}$, $pp \rightarrow \chi_2^0 \chi_1^\pm$ and $pp \rightarrow \chi_1^\pm \chi_1^\pm$ where $\chi^0_1$ and $\chi^0_2$ are respectively Bino and Wino-like. The expected signature of the $\chi_1^\pm \chi_1^\pm$ channel is $\chi^+ \chi^- \rightarrow W^+ W^- + \, \mathrm{MET}$. The $\chi_2^0 \chi_1^\pm$ production channel gives the final state $\chi_2^0 \chi_1^\pm \rightarrow W^\pm\, h\, + \mathrm{MET}$, which should be quite a clear channel at the LHC. Given the dark matter mass of 450 GeV, a 1.5 TeV gluino is expected to be discoverable at the 5$\sigma$ level at LHC-14 given 300 fb$^{-1}$ of data. 

\subsection{Axions}

In addition to WIMPs, axions provide another well-motivated particle candidate for the cosmological dark matter.  The QCD axion provides a possible solution to the Strong CP problem \cite{Peccei:1977hh,Weinberg:1977ma,Wilczek:1977pj}, whereas string axions arise ubiquitously in string / M-theory constructions upon compactification to four dimensions \cite{Svrcek:2006yi,Arvanitaki:2009fg,Acharya:2010zx,Cicoli:2012sz}.  In both situations, axions arise as pseudo-Nambu-Goldstone Bosons (PNGB) of a spontaneously broken $U(1)$ Peccei-Quinn (PQ) symmetry.  The observational consequences of the axion depend primarily on its mass $m_a$, the axion decay constant $f_a$, and the initial misalignment angle $\theta_a$.

For the QCD axion, the mass is related to the decay constant as the axion enjoys a shift symmetry to all orders in perturbation theory and its mass is assumed to result
from non-perturbative QCD instanton corrections  $V(a) = \Lambda_\qcd^4 \left( 1-\cos(a/f^\qcd_a)\right)$, where $\Lambda_\qcd^4 = (m_a f^\qcd_a)^2$, and so there is a 
relation between the mass and decay constant \cite{Peccei:1977hh,Weinberg:1977ma,Wilczek:1977pj},
\be \label{axion_mass}
m_a \simeq  6 \times 10^{-10} \; \mbox{eV} \left( \frac{10^{16} \; \mbox{GeV}}{f^\qcd_a} \right).
\ee
Lab and astrophysical constraints require $f^\qcd_a \gtrsim 10^9$ GeV with stronger model-dependent constraints possible depending on the axion coupling to 
photons and whether the axion is assumed to comprise all of the cosmological dark matter \cite{Agashe:2014kda}.  

Whereas, for string theory axions the relation \eqref{axion_mass} need not apply as the mass can be generated by a number of different non-perturbative effects.  Stated another way, this implies challenging modeling building constraints for realizing a solution to the Strong CP problem in string theory -- notably QCD instantons need to provide the dominant contribution to the low-energy effective mass of the QCD axion.  Yet, investigating the UV completion of the PQ symmetry seems essential given the expectation that global (PQ-like) symmetries are expected to be broken by quantum gravity effects \cite{Holman:1992us,Kallosh:1995hi,Kamionkowski:1992mf}.  Regardless of whether string theory can accommodate the QCD axion these theories do predict PNGBs that can act as all or part of the cosmological dark matter and may have other interesting observational implications.

\subsubsection{Axions from String Theory}
Within string / M-theory constructions, PNGB's naturally arise from the compactification of higher dimensional gauge fields to four dimensions 
\cite{Witten:1984dg,Svrcek:2006yi}.
The higher dimensional gauge invariance of these fields leads to an approximate shift symmetry $a \, \rightarrow \, a \, + \, \, {\rm const.}$ in the $4D$ low-energy effective theory.
This shift symmetry can be lifted by both quantum and non-perturbative effects that depend on the details of the compactification \cite{Svrcek:2006yi}.  
As an example, consider axions arising from compactifications of Type IIB string theory.
To see how axions arise in the theory, we consider the dimensional reduction of the theory to four dimensions by starting from the 
$10$D action in the string frame given by \cite{Polchinski:1998rr}
\be \label{act1}
S_{10}^{IIB}=\frac{1}{(2\pi)^7\alpha^{\prime \; 4}}\int d^{10}x \sqrt{-G} \left[  \frac{1}{g_s^{2}} \left(R[G] - \frac{1}{2} |H_3|^2 \right) + \frac{1}{2} |F_3|^2\right] + \ldots
\ee
where $G_{MN}$ is the ten dimensional string frame metric and $H_3=dB_2$ and $F_3=dC_2$ are the NS-NS and RR three-form fluxes, respectively, with $B_2$ and $C_2$ the corresponding gauge potentials and 
$1/(2 \pi \alpha^\prime)$ is the string tension. The model independent axion $C_0$ and dilaton are combined as the axio-dilaton $\tau =C_0 + i/g_s$ where $g_s = \exp(\phi_0)$ is the string coupling and we will take $C_0$ to be fixed and instead concentrate on the model dependent axions arising from the compactification of the form fields.  The additional terms represented by dots include higher form fields such as $C_4$ (we will concentrate on $C_2$ axions).

The zero modes of $B_2$ and $C_2$ are independent of the co-ordinates of the compact dimensions and can be integrated over chosen two-cycles of the internal geometry giving rise to axions in the four dimensional theory.  To make this explicit, consider compactifying on a Calabi-Yau 3-fold ($CY_3$) and for the form field $C_2$ we make the ansatz \cite{Svrcek:2006yi}
\be \label{axionci}
C_2= \frac{1}{2\pi} c_I(x)\omega^I,
\ee
where the $c_I(x)$ are only functions of the four non-compact space-time dimensions and $I$ labels the two-cycle.  We have introduced the basis forms $\omega^I$ to describe the internal geometry and they obey the normalization condition $\int_{\Sigma_I} \omega^J = (2\pi)^2 \alpha^\prime \delta_I^J$ with the two cycles $\Sigma_I$ giving a basis of the dual homology $H_2(X,\mathbb{Z})$.  The normalization factors of $2\pi$ are chosen for later convenience. Making a similar ansatz for $B_2$ and using this in \eqref{act1} we have
\be
S=\int \frac{d^{10}x }{(2\pi)^7\alpha^{\prime \; 4}} \sqrt{- G} \left[ g_s^{-2}\left(R[G]  - \frac{1}{48\pi^2} G^{nm}G^{lp} \partial_\mu b_{I} \partial^\mu b_{J} \omega^I_{nl} \omega^J_{mp} 
 \right) - \frac{1}{48 \pi^2} G^{nm}G^{lp} \partial_\mu c_{I} \partial^\mu c_{J} \omega^I_{nl} \omega^J_{mp}  \right], 
\ee
where Greek indices run over the four non-compact dimensions and lower-case latin indices denote the compact dimensions.  
 At the classical level, the gauge invariance of the higher dimensional gauge potential implies that the axions can only be derivatively coupled and so we have a shift-symmetric pseudo-scalar in the low energy theory.
This symmetry can be broken in a number of ways, which we will discuss shortly.  
Denoting both types of axions as $a_I$, upon dimensional reduction we find
\be
S_4 = \int d^4x \sqrt{-g} \left( \frac{m_p^2}{2} R[g] - \frac{1}{2} \gamma^{IJ} \partial_\mu a_I \partial^\mu a_J  \right)
\ee
where $g$ is the $4D$ Einstein frame metric and the $4D$ reduced Planck mass is
\be \label{dimreducedplanck}
m_p^2 = \frac{2 {\cal V}}{(2\pi)^7 g_s^2 \alpha^\prime},
\ee
with ${\cal V} =V_6/\alpha^{\prime \; 3}$ the string frame\footnote{ 
Another common convention is to instead work with the Einstein frame volume.
The $10$D string frame is related to the Einstein frame by the 
Weyl rescaling $ G_{MN}^{string} = \exp(\phi/2) G_{MN}^{Einstein}$ and working in units of the string length  
$l_s=2\pi \sqrt{\alpha^\prime}$ the two volumes are related by ${\cal V}_E= (2\pi)^6 {\cal V} /g_s^{3/2}$.} volume of $CY_3$.

For the RR axion the $\gamma^{IJ}$ provide the axion decay constants and depend on the internal geometry as \cite{Svrcek:2006yi}
\be \label{ohmyomega}
\gamma^{IJ}_{\mbox{\tiny RR}}=\frac{1}{6(2\pi)^9 \alpha^{\prime \; 4}} \int \omega_I \wedge \star \, \omega_J,
\ee
whereas for the NS axion one gets the same result multiplied by an extra factor of $g_s^{-2}$.  
Once the specifics of the internal geometry are known one can calculate the $\gamma^{IJ}$ to find the corresponding axion decay constants. This is a non-trivial task, which requires a full specification of the compactification geometry in a Calabi-Yau manifold.  In \cite{Svrcek:2006yi} the $\gamma^{IJ}$ were calculated in a variety of string models assuming compactifications sufficiently symmetric to be amenable to estimates.
\\

{ \it Generating a potential.} Although the higher dimensional gauge invariance guarantees a low energy shift symmetry to all orders in perturbation theory, there are a number of non-perturbative effects that can generate a potential -- e.g. branes wrapped on the corresponding cycle of the axion \cite{McAllister:2008hb,Douglas:2006es} and/or various instanton and condensation effects \cite{Kallosh:1995hi,Becker:1995kb,Dine:1986zy,Dine:1987bq,Acharya:2010zx}. These effects can completely lift the axion from the low energy spectrum or result in a cosine-like potential just as in the case of the QCD axion, albeit at vastly different scales.  In general, the range of axion masses expected from these effects can range over an entire spectrum from very heavy, to string instantons generating masses far smaller than the QCD axion. The large range of model building possibilities has interesting observational implications for many different types of experiments and the resulting theoretical picture has been termed the "axi-verse" \cite{Arvanitaki:2009fg}.
\\

{ \it Moduli Stabilization, the Saxion and a Non-thermal History.}  As familiar from inflationary model building in string theory, it is difficult to introduce a single degree of freedom without examining other ingredients of the fundamental theory.  String / M-theory compactifications come with a number of additional ingredients, which is the cost for working with a UV complete approach.  In particular, one must take into consideration the additional moduli, fluxes, and branes within a chosen framework.  We will not review these difficulties in detail here (see e.g. \cite{Acharya:2010zx,Cicoli:2012sz} and references within).  However, we note that moduli stabilization with 
light axions has a generic challenge as its SUSY partner -- the saxion -- can lead to a number of difficulties.  For example, if the saxion is stabilized supersymmetrically (as Kahler moduli are in KKLT setups before the uplift to deSitter) then the saxion and axion are both expected to have comparable masses.  Thus, if the axion is to remain light and the saxion receives a comparable mass then this can lead to an example of the cosmological moduli problem -- discussed in Section \ref{sec2}.  Whether this is the case or if there can be a large mass splitting is somewhat model dependent.  However, this must be addressed within a given setup to guarantee a successful axion dark matter candidate.  As an example, for a saxion receiving a mass of $\sim 50$-TeV this once again alleviates the cosmological moduli problem and motivates a non-thermal history since the saxion would decay slightly before BBN.  Depending on the initial conditions for the axion this may or may not lead to a successful model for axion dark matter and agreement with the relic dark matter abundance. 

\subsubsection{Axion Dark Matter Density in a Non-thermal History}
Given motivation for axions from both string / M-theory and the Strong CP problem in QCD we now consider the role of the axion as dark matter in a non-thermal universe.
We will not review the standard case of axions in a thermal universe, instead we refer the reader to existing literature \cite{Raffelt:2002zz,Kolb:1990vq,Kim:1986ax}.
However, the standard calculation for the relic density of axion dark matter will hold, even in the non-thermal history with moduli decay, 
if the axion mass is such that axion oscillations begin following moduli decay -- i.e. if $m_a<\Gamma_\sigma$ with $\Gamma_\sigma$ the moduli decay rate given in \eqref{decayrate}. This typically holds
true for axion masses $m_a \lesssim 10^{-15}$ eV, 
and then axion oscillations will begin following moduli decay and the critical density in axions today is \cite{Acharya:2010zx}
\be \label{axion1}
\Omega_a h^2 \simeq 0.1 \; \left( \frac{f_a}{2.0 \times 10^{16} \; \gev} \right)^2 \left( \frac{m_a}{10^{-20} \; \mbox{eV} }\right)^{1/2} \langle \theta_0^2 \rangle
\ee
where $\Omega_a \equiv \rho_a / \rho_c$ is the density of axions today normalized to the critical density and this expression 
is accurate up to $\mathcal{O}(1)$ corrections due to effects such as anharmonicity \cite{Fox:2004kb}. We note that for the special case of a QCD axion \eqref{axion1} can be further simplified by making use of \eqref{axion_mass}.  As mentioned above, the result \eqref{axion1} depends on the axion mass and decay constant, as well
as the misalignment angle of the axion from its minimum $\theta_0$.  In most of the literature the initial value $\theta_0$ is taken as random variable with an average squared value $\langle \theta_0^2 \rangle \sim \pi^2 / 3$.  However, the actual value depends on a number of assumptions, including the status of the PQ symmetry during inflation \cite{Dine:2004cq}.  In this review we will leave this value unspecified to be as general as possible, invoking the standard average when numeric answers are required.  We see from \eqref{axion1} that for a decay constant in the vicinity of the GUT or string scale ($f_a \simeq 10^{16}$ GeV) we must either have extremely light axions $m_a \ll 10^{-20}$, or fine-tune the initial misalignment angle $\theta_0 \ll 1$, in order not to {\it over-close the universe}, i.e. to be consistent with cosmological observations which imply the abundance of axions today must be $\Omega_a h^2 \lesssim 0.1$.
\\

\noindent
{\it Axion oscillations before moduli decay.} Next we consider when the axions begin oscillating prior to moduli decay requiring $m_a \gtrsim 10^{-14}$ eV.  
In this case the moduli decay and the entropy produced will dilute the relic abundance as first discussed in \cite{Lazarides:1990xp}.
For completeness, we now briefly review this calculation of the relic abundance as it appears in \cite{Fox:2004kb}.

The equation of motion for the axion in the background of the oscillating moduli is given by
\be
\ddot{a}+3H\dot{a}+m_a^2a=0,
\ee
where $H\sim 2/(3t)$ in the effectively matter dominated epoch and  $m_a(T)$ is the axion mass which can now depend on the temperature \cite{Fox:2004kb}.
The energy density of the axion is $\rho_a=(\dot{a}^2 +  m_a^2 a^2)/2$ and once coherent oscillations begin (typically in much less than a Hubble time, $ \Delta t_H \sim H^{-1}$)
then the average energy density evolves as $\langle \rho_a \rangle = \langle \dot{\theta}^2 \rangle = \langle m^2(T) \theta^2 \rangle$ and the pressure then vanishes.
We then have
\be
\dot{\rho}_a=-\left( 3H - \frac{\dot{m_a}}{m_a} \right) \rho_a,
\ee
where the time dependence of the mass follows from the temperature dependence and so the energy density evolves as
\be
\rho_a= \rho_0 \frac{m_a(T) }{a(t)^3},
\ee
where $a(t)$ is the scale factor (not to be confused with the axion).
The temperature dependence of the mass implies that the energy density does not scale like matter but that the number density ($n_a$) does.
Thus, if the axions start to oscillate during the moduli phase their number density compared to that of the moduli (which we recall also scale as $\sim 1/a(t)^3$) is constant 
\be \label{e4}
\left. \frac{n_a}{n_\sigma} \right\vert_{t=t_{osc}}= \frac{ \frac{1}{2} m_a(T_{\ttt{osc}}) f_a^2 \Delta \Theta_{eff}^2 }{3 H^2_{\ttt{osc}} m_p^2 / m_\sigma},
\ee 
where $H_{\ttt{osc}}$ and $T_{\ttt{osc}}$ denote the Hubble parameter and temperature at the onset of {\it axion} oscillations $t=t_{osc}$ and $\Delta \Theta_{eff}^2 \equiv f(\langle \theta_0^2 \rangle) $ is the axion displacement including various finite temperature and other corrections which can be found as a function
of the zero temperature displacement $\theta_0$ \cite{Fox:2004kb}.
The amount of axions after moduli decay and at the time of reheating $t_r$ is 
\bea 
\left. \frac{n_a}{s} \right\vert_{t>tr} &=& \frac{n_a}{\rho_\sigma} \left( \frac{\rho_\sigma}{s(T_r)}\right).
\eea
where $s(T)$ is the entropy density and we assume that the moduli do not decay into axions (we will discuss other cases soon).
Comoving energy is conserved during the decay so for instant reheating we can relate the moduli density to the radiation after reheating and $\rho_\sigma=\rho_r=\pi^2 g_\ast T_r^4/30$, so that
\bea \label{axion_eqn}
\left. \frac{n_a}{s} \right\vert_{t>tr} &=& \frac{n_a}{\rho_X} \left( \frac{\rho_r}{s(T_r)}\right), \nonumber \\
&=& \frac{n_a}{\rho_X} \left( \frac{3}{4} T_r \right),
\eea
where we used $s(T_r)=2\pi^2 g_s(T_r) T_r^3/45$ and $g_s(T_r) \simeq g_\ast(T_r)$.
Using that \eqref{axion_eqn} is conserved until today, the critical density is
\be
\Omega_a = \left(\frac{m_a s_0}{\rho_c} \right) \left. \frac{n_a}{s} \right\vert_{t>tr},
\ee
where $s_0$ and $\rho_c=3H_0^2m_p^2$ are the entropy and critical density today, respectively.
Using $\rho_c / s_0 = 3.7 \times 10^{-9} h^2$ GeV with $h$ the Hubble parameter in units of $100$ km/s/Mpc and \eqref{e4} and \eqref{axion_eqn} we have
\bea
\Omega_a h^2 &=& 3.0 \times 10^{8} \; \left( \frac{f_a^2 \Delta \Theta^2 T_r}{\xi_{\ttt{osc}} m_p^2} \right), \nonumber \\
&=& 5.3  \;  \left( \frac{f_a}{10^{16} \gev} \right)^2 \left( \frac{T_r}{1 \; \mev} \right) \xi_{\ttt{osc}}^{-1} \; \Delta\Theta_{eff}^2 \; \;\;\;
\label{abundance}
\eea 
where $\xi_{\ttt{osc}} \equiv m_a(T_{\ttt{osc}}) / m_a(0)$ is the ratio of the finite temperature mass to the zero temperature mass and is typically an order one number \cite{Fox:2004kb}. 
This result is interesting in that as long as $m_a \gtrsim 10^{-14}$ eV the resulting abundance is independent of the mass of the axion.  
We see from \eqref{abundance}, for low reheat temperatures and axion decay constants near the GUT scale $f_a \simeq 10^{16}$ GeV,  
the initial displacement angle would require some level of fine-tuning ($\Delta\Theta_{eff} \ll 1$) in order to be consistent with the observed dark matter abundance.  
This presents a challenge for many string and M-theory based setups containing axions.  For example, in \cite{Acharya:2010zx} it was found that for M-theory approaches with gauge coupling unification that the masses of the axions would be distributed logarithmically between $10^{-33}$ eV and about an eV, but that typically $f_a$ should be near the GUT scale.  However, it is again important to stress that a knowledge of the nature of the PQ symmetry during and after inflation is important for establishing the likely values of the displacement \cite{Dine:2004cq}.  

\subsubsection{Observational Constraints on Axions}
In addition to bounds on axions from the dark matter relic density, their behavior both during inflation and at the time of moduli decay can provide important constraints.
For axions with masses comparable or lighter than the Hubble scale ($H_I$) during inflation $m_a \lesssim H_I$ axions can generate an isocurvature component in the primordial perturbations. Whereas, axions which are relativistic near the epoch of BBN can act as an additional source of dark radiation which is also constrained by both CMB observations and BBN.  Both of these constraints can be quite stringent and depend on assumptions of the cosmic history.  We will discuss both of these constraints and their connection with the moduli phase in Section \ref{cmb_constraints}.  

Many of the constraints on axions are rather independent of whether the post-inflationary history is thermal or non-thermal. 
We will mention these briefly and refer the interested reader to the literature for more details -- see e.g. \cite{Arvanitaki:2009fg} and references within.
The relevant constraints depend primarily on the mass range of the axions and on the strength of the coupling to photons
$\sim a \, {\bf E}\cdot {\bf B}$.  In \cite{Arvanitaki:2009fg}, it was shown axions with masses $m_a \lesssim 10^{-28}$ eV (and with appreciable electromagnetic coupling) lead to rotation of the CMB and the level of constraint will depend on the strength of the coupling -- of course for axion dark matter candidates this coupling must be quite weak.  For axions in the mass range $10^{-28}$ eV $< m_a \lesssim 10^{-18}$ eV step-like features in the matter power spectrum can become important for structure formation on small scales.  For larger masses in the range $10^{-18}$ eV $<m_a < 10^{-10}$ eV axions can form bound states with black holes leading to a spin-down of the black hole and a mass gap in the spectrum.  Finally, for axions with masses larger than around $10^{-10}$ eV and with a significant coupling to photons, axion-photon conversion can occur in strong magnetic fields resulting in signals both in galaxies and in more compact objects such as pulsars.  

\section{Baryogenesis and the Cosmic Coincidence \label{baryosection}}

String cosmology has been most widely studied in the contexts of inflationary, axionic, and dark matter physics. On the other hand, the baryon asymmetry typically relies on model-building in the visible sector. While realistic vacua that approximate the Standard Model have been constructed in different corners of M-theory, it is safe to say that extensions required for baryon asymmetry have not been a model-building priority. This is understandable, since processes such as electroweak baryogenesis \cite{Cohen:1993nk} or leptogenesis \cite{Fukugita:1986hr} are relatively independent of a particular UV completion, according to the common lore. 

There is every reason to expect that the viability of canonical processes like electroweak baryogenesis and leptogenesis does, in fact, have an unavoidable dependence on UV physics. The decay of moduli will dilute any previous population of baryons. Since electroweak baryogenesis depends on processes at scales $\sim \mathcal{O}(100)$ GeV, and moduli decay at scales $\sim \mathcal{O}(1)$ GeV, this dilution factor can and should be large.

There are then two options. The first is to rely on a high scale baryogenesis process such as Affleck-Dine (AD) baryogenesis which produces a large baryon asymmetry that can withstand subsequent dilution. The second is to implement baryogenesis at a post-sphaleron stage. We discuss these options in turn.

\subsection{AD Baryogenesis: UV Constraints and Challenges}

Affleck-Dine (AD) baryogenesis \cite{Dine:1995kz}, \cite{Dine:1995uk}, \cite{Affleck:1984fy} relies on very general properties of supersymmetric theories: the existence of flat directions. An inflationary sector produces coherent oscillations along a supersymmetric flat direction $S$, and the interaction between the inflationary sector and the flat direction leads to dynamics in field space that ultimately breaks CP and baryon number. This interaction occurs through Planck-suppressed operators. The setting is effective $D=4, N=1$ supergravity, and string phenomenology plays a definite role.

The relevant potential for the AD field $S$ in the early Universe is \cite{Dine:1995uk,Dine:1995kz}, as explained below,
\begin{equation}
V(S)=(-c_HH^{2}+m_{S}^{2})|S|^{2} + \left( \frac{aH+Am_{3/2}}{M^{n-3}%
}\lambda S ^{n}+h.c.\right) +|\lambda |^{2}\frac{|S|^{2n-2}}{M^{2n-6}}%
,\label{vS}
\end{equation}%
where $c_H$ and  $a$ correspond to a Hubble-induced mass and $A$-term, whereas $m_S$ and $A$ are the soft-SUSY breaking mass and $A$-term, respectively. The last term in the potential corresponds to a higher dimensional operator in the superpotential, $W \supset \lambda S^{n}/M^{n-3}$ ($M$ is the cutoff scale where new physics appears).

The curvature along $S$ is dominated by the Hubble-induced mass term for $H \gg m_{\rm soft}$. If $c_H > 0$, the field sits at the origin. However, if $c_H < 0$, the minimum is non-zero and the field tracks this minimum until $H \sim m_S$. The field settles into one of the $n$ discrete vacua given by the Hubble induced $A$-term. When $H \sim m_{S}$, the new minimum develops at the origin and the field begins to oscillate around it. The soft $A$-term becomes dominant over the Hubble-induced $A$-term, and the field acquires an angular motion to settle into a new phase. The baryon number violation becomes maximal at this time and imparts an asymmetry to the condensate.

The different components of \eqref{vS} are non-trivial to justify from a top-down perspective. We describe these challenges in turn. 
\\

\noindent
$(i)$ {\bf Negative Hubble-induced mass.} The foremost requirement is a tachyonic Hubble-induced mass term for the field $S$. This requires a condition on the geometry of the moduli space. The Kahler potential and superpotential can be written as
\bea \label{KW}
K &=& \widehat{K}(\sigma_i,\overline{\sigma}_i) + \widetilde{K}_{\alpha \overline{\beta}}(\sigma_i,\overline{\sigma}_i) \overline{S}^{\alpha}  S^{\beta}+ \ldots \nonumber \\
W &=& \widehat{W}(\sigma_i) + \frac{1}{6}Y_{\alpha \beta \gamma} S^{\alpha \beta \gamma} \,\,,
\eea
where $S$ denotes a generic visible sector field. The soft mass is given by 
\be \label{formulaformasses}
m_{S \Sbar}^2 \,\, = \,\, V_0(1 + \mathbb{H}[S,\sigma]) \, + \, 3m_{3/2}^2(\frac{1}{3} + \mathbb{H}[S,\sigma])\,\,,
\ee
with the holomorphic bisectional curvature $\mathbb{H}[S,\sigma])$ being given by 
\be
\mathbb{H}[S, \sigma] \, = \, - \frac{R_{\sigma \bar{\sigma} S \Sbar}}{g_{\sigma \bar{\sigma}} g_{S \Sbar}}\,\,.
\ee
where $R$ denotes the Riemann curvature tensor of the moduli space.

Since during inflation we can take $V_0 \gg m_{3/2}$, the requirement of a tachyonic Hubble-induced mass reduces to
\be
\mathbb{H}[S,\sigma] \, < \, -1 \,\,.
\ee
This condition has been explored in a wide variety of contexts within type IIB string theory \cite{Dutta:2010sg, Dutta:2012mw}. For a non-local modulus $\sigma$, it was found that the Hubble-induced masses are generally positive, with 
\be
c_H \, = \, 2/3 \,\,\,,
\ee
making AD baryogenesis difficult to satisfy. If the modulus $\sigma$ corresponds to deformations of the local cycle on which the visible sector is constructed, on the other hand, then the induced masses are also typically positive, although for suitable choices of $\widetilde{K}_{\alpha \overline{\beta}}(\sigma_i,\overline{\sigma}_i)$ a tachyonic mass can be generated. If the visible sector construction is at a singularity, then the
induced mass depends on appropriate choice of fluxes on the mirror type IIA manifold.
Aspects of this problem have recently been discussed in \cite{Garcia:2013bha, Marsh:2011ud, Kamada:2012bk, Higaki:2014eda}.
\\

\noindent
$(ii)$ {\bf Hubble-induced A-terms.} To obtain successful AD bayogenesis, the presence of a Hubble-induced $A$-term is necessary. The $A$-term is given by 
\be \label{atermkl}
A_{\alpha \beta \gamma} \, = \, F^{i} \left( \partial_i Y_{\alpha \beta \gamma} + \frac{1}{2}\widehat{K}_i Y_{\alpha \beta \gamma} - \Gamma ^{\gamma}_{i( \alpha} Y_{\beta \lambda ) \gamma} \right) \,\,,
\ee
where $\Gamma$ is the connection on moduli space, and the $i$-th modulus $\sigma_i$ acquires the dominant $F$-term during inflation (we will simply call it $\sigma$ henceforth). 

We need the presence of a Yukawa coupling $Y_{\alpha \beta \gamma}$ in the superpotential in \eqref{KW}. 
To obtain Hubble-induced $A-$terms during inflation, the various contributions in (\ref{atermkl}) should acquire a non-zero value. Generally, the dependence of the Yukawa coupling on the modulus sector depends on the details of the local model building. Consider a visible sector localized near a small four-cycle $\tau_s \, = \, {\rm Re}\, T_s$. Due to holomorphy and the shift symmetry of the Kahler moduli, they cannot appear at any level in perturbation theory in $W$, and hence, importantly, in $Y$ of (\ref{KW}). The entire dependence on Kahler moduli appears in the normalized Yukawa couplings, given by
\be \label{YK}
\widehat{Y}_{\alpha \beta \gamma}(\tau_s, \mathcal{U}) \, = \, e^{\widehat{K}/2}\frac{Y_{\alpha \beta \gamma}(\mathcal{U})}{\left( \widetilde{K}_{\alpha} \widetilde{K}_{\beta} \widetilde{K}_{\gamma}\right)^{\frac{1}{2}}} \,\,.
\ee
This should only depend on local geometric data $\tau_s$ and complex structure moduli $\mathcal{U}$, but not the overall volume or other Kahler moduli.

The first term in (\ref{atermkl}) requires that the Yukawa couplings depend on the modulus ($\sigma$) that acquires the dominant $F-$term during inflation. This term vanishes in general since the Kahler modulus corresponding to deformations of the cycle on which the visible sector is constructed is typically not the dominant source of supersymmetry breaking during inflation. Obtaining a non-zero value for the other two terms in (\ref{atermkl}) depends both on the geometry as well as the specific details of inflation. A detailed study of these issues is merited.

The AD baryogenesis scenario also has a number of correlated predictions coming from the presence of a light scalar field during the inflationary era. One important issue is the back-reaction on the inflaton. The AD field ($S$) makes a transition from the origin to its non-zero vacuum configuration in field space when the Hubble-induced tachyonic mass term becomes dominant. Depending on when this transition takes place, there will be back-reaction on the inflationary potential, leading to features on the power spectrum, which will be constrained by data \cite{Marsh:2011ud}. The presence of a scalar field coupling to the inflationary sector should also lead to constraints from non-Gaussianity. Both of these issues deserve further study.

\subsection{Moduli and the Cosmic Coincidence Problem}
Cosmological observations not only determine precisely the relic abundance of dark matter and baryons, but also imply an interesting connection between their relative amounts $ \Omega_{dm} /\Omega_B\approx 5$ leading to what some have called the `Cosmic Coincidence Problem'.  In \cite{Kane:2011ih} it was argued that moduli decay in non-thermal histories might address this `coincidence' -- the primary feature being that both dark matter and baryons can be produced from the decay of the lightest modulus.
Keeping in mind that moduli decay will dilute any initial abundance of particles, it is advantageous in non-thermal histories that AD baryogenesis typically over-produces baryons.  
Indeed in \cite{Campbell:1998yi} the authors pointed out that the late decay of a Polonyi field could act to dilute the baryon asymmetry to an acceptable level.  More recently, 
in \cite{Kane:2011ih} it was noted that because dark matter can also be produced in the decay, this creates a subtle connection between the two relic densities that could account for the baryon-dark matter coincidence problem.
To understand the connection it is useful to consider the ratio of the baryon to moduli co-moving densities
\begin{equation}
\frac{Y_{B}^{0}}{Y_{\sigma}^{0}}\simeq \left( \frac{m_{\sigma}}{m_{S }}\right)
\left( \frac{S_{0}}{\sigma_{0}}\right) ^{2} \left( \frac{n_{B}}{n_{S}}\right) _{i}, \label{ratio}
\end{equation}%
where $\sigma_{0}$ is the amplitude at the start of the moduli oscillation, and the subscript $i$ and superscript $0$ denote the values before and after modulus
decay, respectively.

To obtain the baryon number density after decay, the dilution factor $\Delta=s_{after}/s_{before}$ must be used. We obtain
\begin{equation} \label{baryonfinal}
Y_{B}^{0} \rightarrow Y_B=\frac{Y_{B}^{0}}{\Delta} = \frac{n_B}{s_{\text{after}}} \simeq \frac{n_{\sigma}}{s_{\text{after}}}\left(\frac{Y_B^0}{Y_{\sigma}^0}\right) \simeq \frac{3}{4}\frac{T_{r}}{m_{S}}\left( \frac{S_{0}}{\sigma_{0}}\right) ^{2} \left( \frac{n_{B}}{n_{S}}\right) _{i},
\end{equation}%
which demonstrates that the baryon abundance in this approach is intimately related to the ratio of the initial amplitudes of the AD field and the modulus.
The AD fields displacement depends nontrivially on the dimension of the non-renormalizable operator that lifts the flat direction. Since larger $n$ leads to larger displacement, and therefore larger contribution to the baryon asymmetry, one can focus on the flattest directions in MSSM that require the largest $n$ to get lifted\footnote{We assume all non-normalizable operators that are allowed by gauge invariance and R-parity are generated.}. As showed in Ref.~{\cite{Gherghetta:1995dv}}, the flattest direction (one of the $Q, u, e$ combinations) corresponds to $n=9$. 
Assuming that the non-renormalizable operator is generated at the reduced Planck scale $M\sim M_p$ 
and taking $m_{S}\sim 10^5$~GeV we find $S_0\sim 10^{16}$~GeV. 
For the next flattest direction (one of the $d,L$ combinations) -- which is not lifted until $n=7$ --
we have $S_0\sim 3\times 10^{15}$~GeV. So we can see that these directions in the MSSM naturally 
have amplitudes two or three order of magnitudes smaller than $M_p$, and it follows that the correct baryon asymmetry
 can be obtained for $T_r \sim 100 $ MeV, $m_S \sim 100$ TeV, and $(S_0 / \sigma_0) \sim 10^{-2}$. 

The modulus will also decay and provide dark matter, and the relic density will depend on whether the number density produced is above ('annihilation scenario') or below ('branching scenario') the critical number density. In either case, the final baryon-dark matter relic density ratio will depend on the factors in \eqref{baryonfinal} and the annihilation cross section of the dark matter or the branching ratio of $\sigma$ into dark matter. By a reasonable choice of parameters, one would obtain $\Omega_{B}/\Omega_{DM} \sim 1:5$.

We note that similar approaches to the baryon-dark matter coincidence problem that depend on moduli physics have been undertaken recently \cite{Kawasaki:2015cla, Doddato:2012ja, Cheung:2011if, Bell:2011tn, Shoemaker:2009kg, Roszkowski:2006kw}.  A more direct solution to the coincidence problem would result if the {\it both} the baryons and the dark matter resulted from the decay of the modulus.  This idea was first proposed in \cite{Kitano:2008tk}, and it would be very interesting to see if this approach could be UV completed in String or M-theory.

\subsection{Baryogenesis at Late Times}
Another approach to baryogenesis in non-thermal histories can be implemented by a leptogenesis-like mechanism resulting from modulus decay.
This approach invokes additional model building, but it implies an intriguing connection between the need for $Y_{\sigma} \, = \, (3 T_r/ 4 m_{\sigma}) \, \sim \, 10^{-8}$ to be a small number and that baryogenesis requires small number densities $\sim \, \mathcal{O}(10^{-10})$. This small value can be driven by the smallness of $Y_\sigma$, with the remainder factor of $10^{-2}$ coming from branching ratios and loop factors \cite{Allahverdi:2010rh,Allahverdi:2010im}. 

As an explicit model, we can consider the following extension of the MSSM, containing two flavors of singlets $N_{\alpha}$ and a single flavor of colored triplets $X, \overline{X}$ (with hypercharges $+4/3,-4/3$ respectively). The superpotential is given by
\bea \label{superpot2}
W_{\rm extra} = \lambda_{i\alpha} N_{\alpha} u^c_i X + \lambda^\prime_{ij} d^c_id^c_j \overline{X}
+ {M_{\alpha} \over 2} N_{\alpha} N_{\alpha} + M_{X} X \overline{X} \, . \nonumber \\
\eea
The interference between the tree-level and one-loop diagrams in the $N_\alpha \rightarrow X u^c$ decay generates a baryon asymmetry, in a manner reminiscent of leptogenesis. The asymmetry per decay $\epsilon$ is loop suppressed and naturally $\epsilon \sim \mathcal{O}(0.1)$ for $\mathcal{O}(1)$ phases and couplings. Taking $Y_\sigma \sim 10^{-7}$, the correct baryon asymmetry can be obtained for a branching of the modulus into  the species $N$ given by $Br_{N} \sim 10^{-2}$. From counting of the degrees of freedom this branching ratio appears to be somewhat natural.

The modulus will generally produce dark matter $\chi$ in addition to the the field $N$ required for baryogenesis. 
Considering that the number density of the dark matter produced is below the critical density required for further annihilation, one has for the baryon and dark matter density ratio
\be
\frac{\Omega_{\rm B}}{\Omega_{\rm DM}} \, \simeq \,
\frac{1 ~ {\rm GeV}}{m_\chi} \times \frac{\epsilon \, {\rm Br}_N}{{\rm Br}_\chi} \,\,.
\ee
The branching scenario requires ${\rm Br}_\chi \lsim 10^{-3}$. One then obtains the correct coincidence for $m_\chi \, \sim \, \mathcal{O}(10)$ GeV, but other values of $m_{\chi}$ can also be accommodated.

\section{Gravitational Constraints: CMB and Large Scale Structure} \label{cmblssDM}
In this section we consider possible implications of a non-thermal history on the CMB, BBN and the Matter Power Spectrum.
Thus far, many of the observational consequences of a non-thermal history have relied on a connection with interactions or decays to Standard Model particles.  
In the case of dark matter, we saw that non-thermal WIMPs can have unexpected properties (e.g. a stronger self annihilation cross section than the thermal case) and as a result we saw that (in)direct detection of dark matter experiments could help constrain the reheat temperature.  However, these probes rely on the assumption that dark matter is composed of WIMPs and have a connection to electroweak-scale interactions. 

{\it What if dark matter is not directly associated with weak-scale physics?} We discussed one example -- axions -- and we saw that if these particles are light they could prove difficult to detect. If part of the cosmological dark matter has negligible interactions with (or decays to) the Standard Model, this would present a substantial challenge to identifying its particle nature.  Moreover, as LHC prepares to probe energies in excess of $10$ TeV, there is, as yet, no evidence for physics beyond the Standard Model.  If nature has decides to play a cruel trick on us and the scale of new physics lies significantly above the electroweak scale, our only near-term possibility for probing new physics may rely on cosmological observations. In this section we present implications of the non-thermal history resulting from string / M-theory approaches on gravitational and cosmological based experiments. In particular, we consider what the early matter dominated phase could imply for the formation of primordial structures (such as mini-halos and primordial black holes), we also discuss implications for the matter power spectrum, and different ways in which the matter dominated phase can influence observations of the CMB.  

\subsection{Implications for the Matter Power Spectrum}
\subsubsection{Enhanced Structure Growth on Small Scales}
\begin{figure}[t!]
\begin{center}
\includegraphics[scale=0.64]{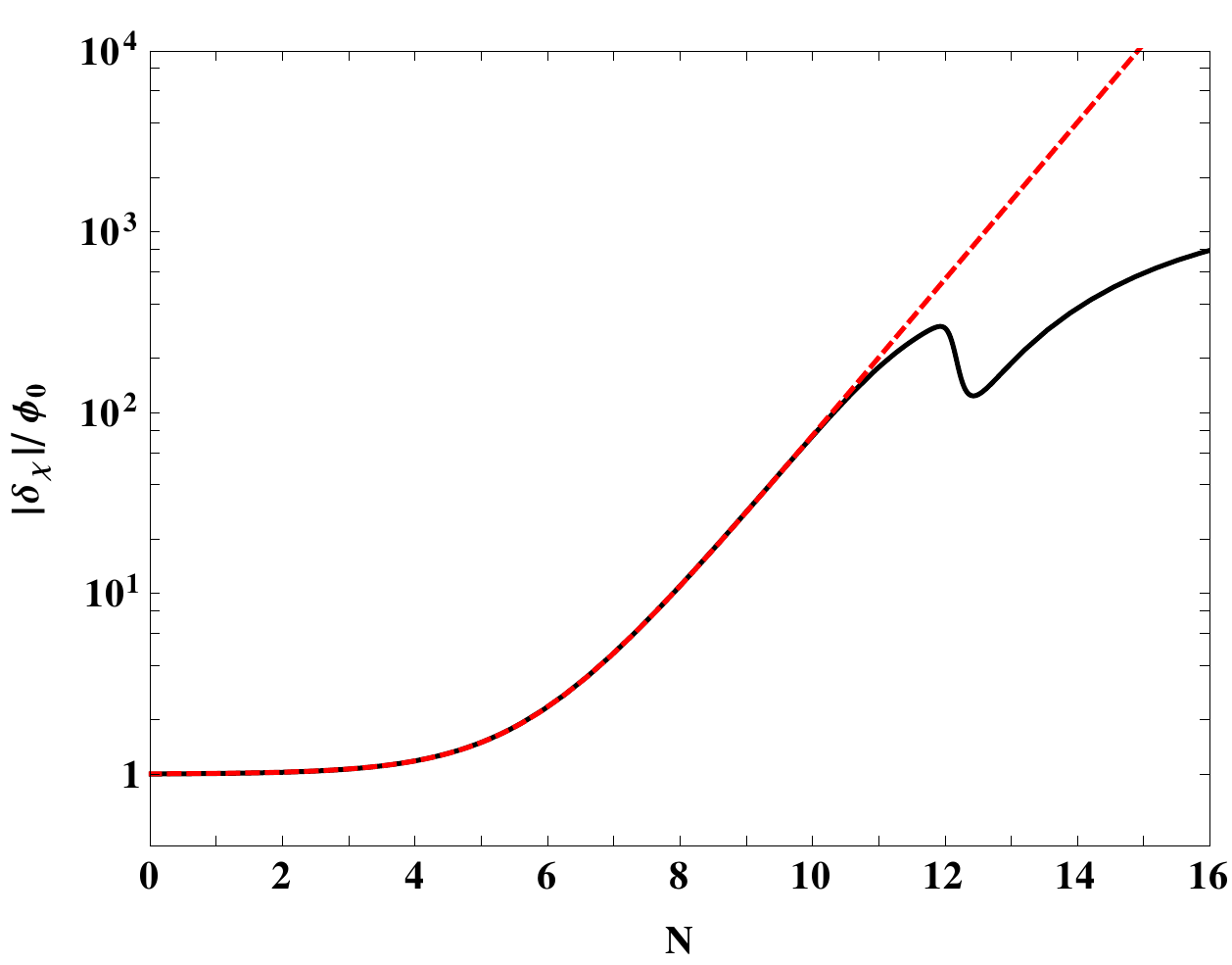}\includegraphics[scale=0.65]{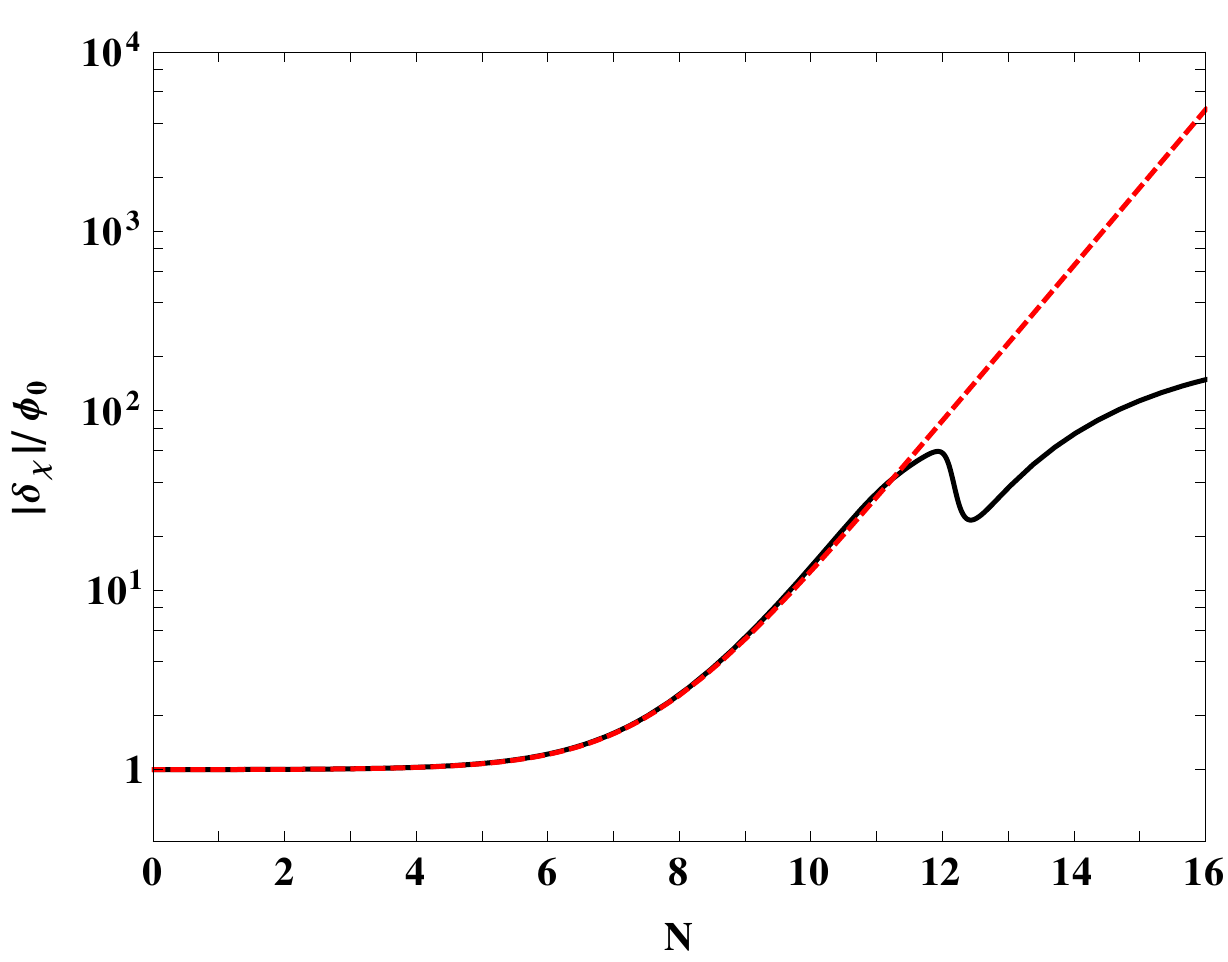}
\hspace{0.5cm}
\end{center}
\caption{Evolution of (normalized) dark matter contrast ($\delta_{dm} \equiv \delta \rho_{dm} / \rho_{dm}$) in a non-thermal cosmology for modes entering the horizon at $N\simeq 5$ (Left) and $N\simeq7$ (Right) where $\Phi_0$ is the initial metric perturbation. As the mode enters the horizon it grows linearly with the scale factor and modes entering earlier experience more growth.  After the universe becomes radiation dominated at $N\simeq 12$, the amplitude of the density contrast decreases due to rapid annihilations of dark matter particles, and then the density grows logarithmically as expected. \label{fig_dm}}
\end{figure}
\begin{figure}[t!]
\begin{center}
\includegraphics[scale=0.580]{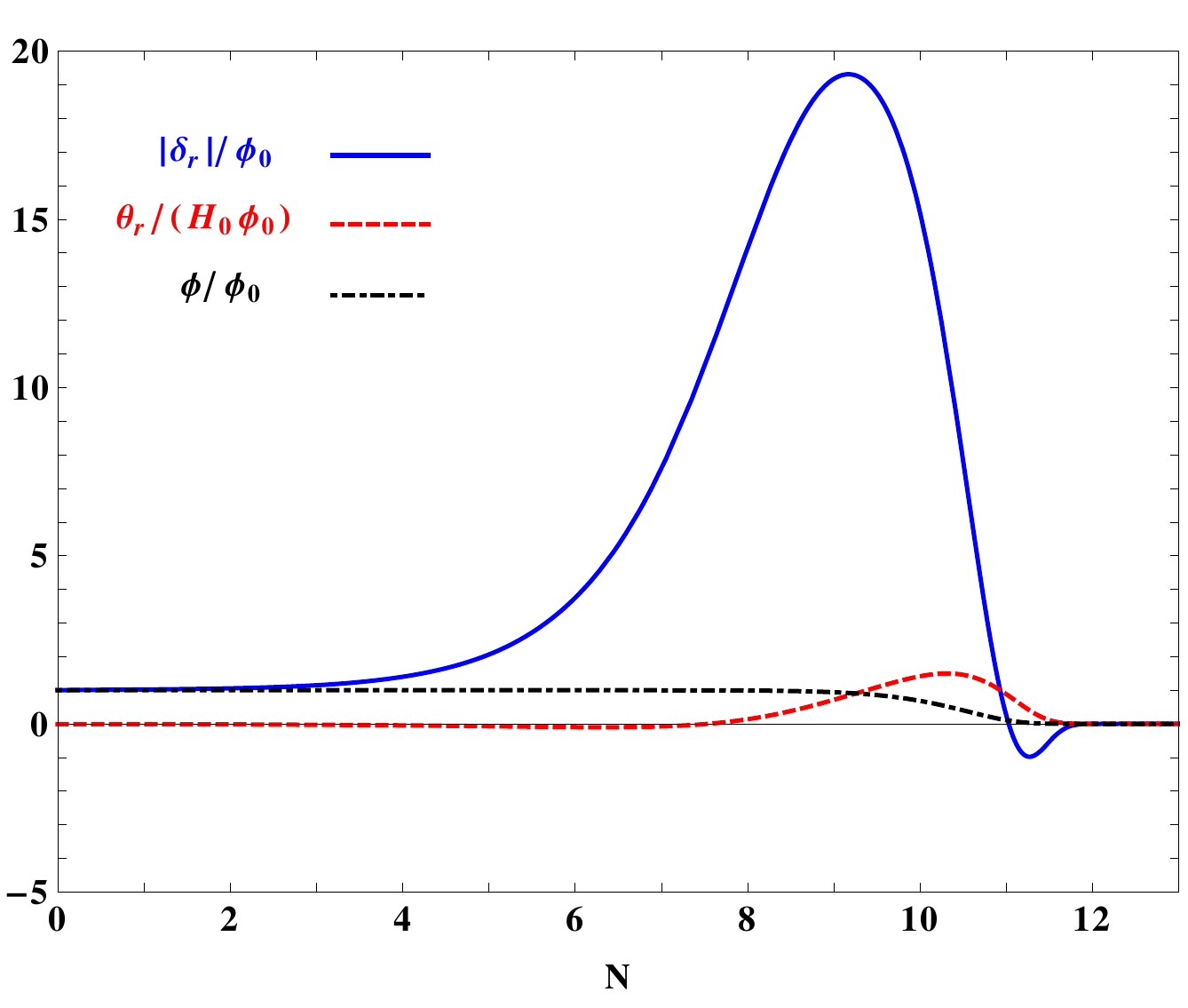}
\includegraphics[scale=0.595]{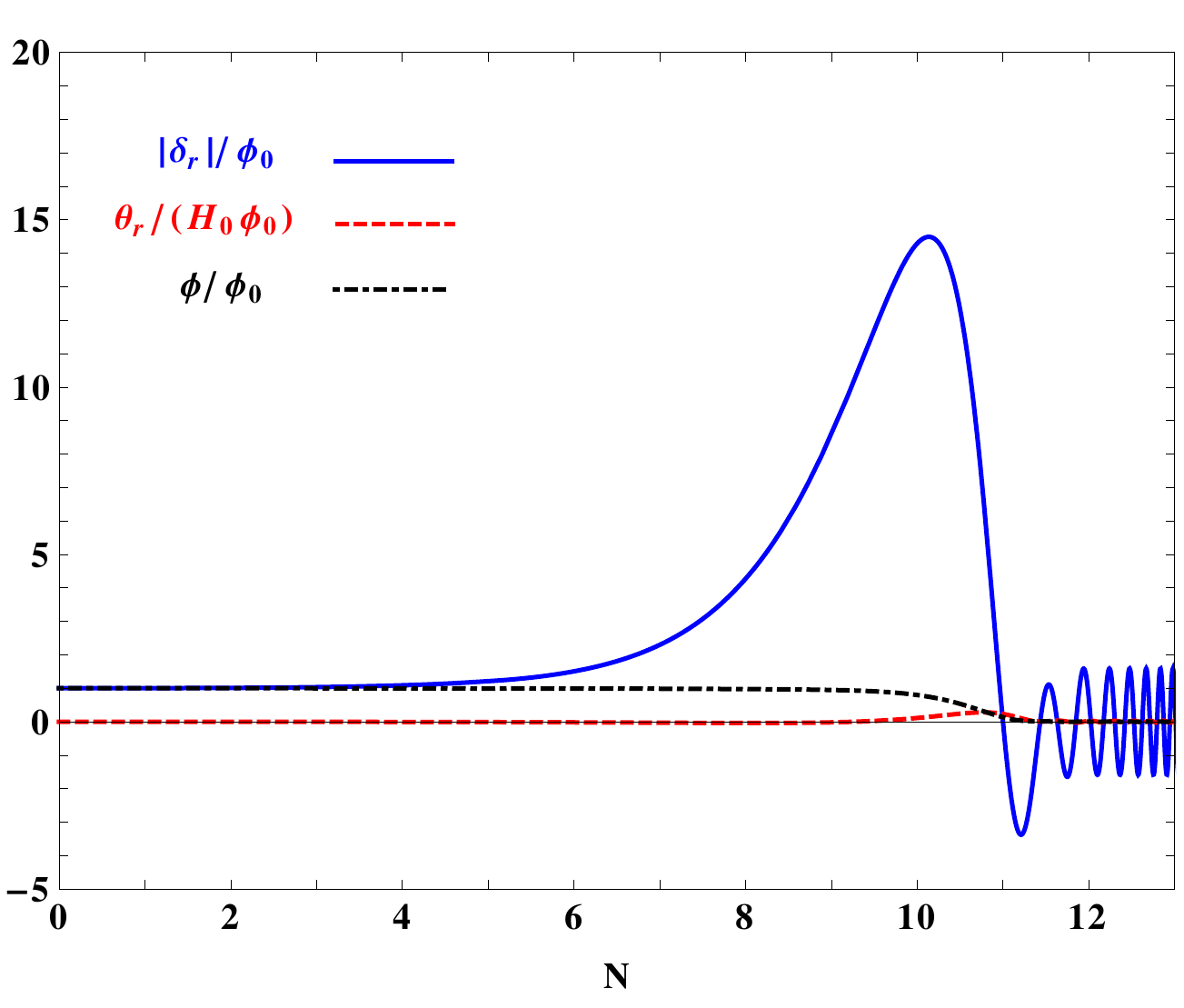}
\hspace{0.4cm}
\end{center}
\caption{Evolution of the (normalized) radiation density contrast $\delta_r \equiv \delta \rho_r / \rho_r$ and velocity perturbation $\theta_r$ for the modes $k/H_0=0.1$~(Left), $k/H_0=0.04$~(Right) where $\Phi_0$ is the initial metric perturbation. This mode crosses the horizon at $N \simeq 4.5$~(Left), $N \simeq 6.5$~(Right). 
In this non-thermal cosmology the universe is effectively matter dominated until $N \simeq 10.5$ e-foldings after which the universe becomes radiation dominated. \label{fig2}}
\end{figure}
In conventional models for structure formation growth of structure only becomes significant following matter-radiation equality.  This is because dark matter perturbations $\delta \rho / \rho$ that enter the horizon only grow logarithmically with the scale factor in a radiation dominated universe $\delta \rho / \rho \sim \log a(t) \sim \log t$, whereas they grow linearly in a matter dominated universe $\delta \rho / \rho \sim a(t) \sim t^{2/3}$.  However, in non-thermal cosmologies we have seen that the universe is matter dominated prior to BBN.  This suggests an early period of growth and a new scale at smaller wavelengths in the matter power spectrum.  This possibility was first studied in the absence of annihilations in \cite{Erickcek:2011us} and later more generally in \cite{Fan:2014zua}.  The results are summarized in Figure \ref{fig_dm}. It was found that the moduli phase would introduce a new period of growth for dark matter perturbations and that the size of the horizon at reheating could provide a new scale for determining the size of the smallest possible dark matter substructures.  Indeed, for reheat temperatures near the limit set by BBN it was shown that a detectable level of Earth-sized, ultra-compact mini-halos may result.  Of course, whether such structures can survive depends on a number of details, including whether free-streaming and kinetic decoupling effects would destroy the initial formation process. In \cite{Fan:2014zua}, it was found that for typical WIMP candidates in string-based non-thermal cosmologies that free-streaming and kinetic decoupling would instead determine the smallest structures.  However, as discussed in \cite{Erickcek:2011us,Fan:2014zua}, there are exceptions pointing to new directions for model building and a possible signature of the non-thermal period.  One example is if the WIMPs are produced non-relativistically in the moduli decays (so that free-streaming is negligible) and if these particles become kinetically decoupled before reheating.  Whether such models can be realized from fundamental theory remains an open question, although if such structures do survive they could lead to new expectations for indirect detection experiments.

In addition to enhanced growth of dark matter perturbations, it was also found in \cite{Erickcek:2011us,Fan:2014zua} that there can be interesting effects on radiation perturbations. These perturbations also grow during the moduli dominated phase, and then following reheating those that entered the horizon will oscillated with an amplitude suppressed relative to the situation in a strictly thermal history.  As seen in Figure \ref{fig2}, the suppression in the radiation perturbations arises at the peak of scalar decay when most of the radiation is created by both dark matter annihilations and decays of the moduli.  The suppression of the radiation perturbations can lead to a new critical scale for dark matter structures if dark matter is produced thermally following reheating \cite{Erickcek:2011us,Fan:2014zua}, since the suppression would prevent acoustic oscillations on that scale from wiping out structure \cite{Loeb:2005pm}. However, again it was argued in \cite{Fan:2014zua} that this would require non-standard WIMPs, since for the interesting case of low reheat temperatures ($\sim$ MeV) this temperature would typically be far below the thermal freeze-out temperature ($\sim$ GeV).
  
In summary, it was found in \cite{Erickcek:2011us,Fan:2014zua} that non-thermal histories do not face any challenges in regards to disrupting structure formation.  Moreover, if we consider less typical SUSY WIMP dark matter candidates it way be possible in those models to distinguish between non-thermal and thermal scenarios through future observations. 

\subsubsection{Primordial Blackhole Constraints}
Above we primarily focused on the growth of dark matter and radiation perturbations during the moduli phase.  However, both of these components are typically sub-dominant in string-based approaches and instead the moduli significantly dominate the energy density.  We argued above that the transfer of these perturbations into dark matter and radiation can have interesting effects in some cases.  But one can ask if the growth of moduli perturbations during the early matter domination can lead to interesting constraints?  In particular, it was realized many years ago that an early matter dominated (dust) phase could lead to a significant production of primordial black holes (PBHs) \cite{Polnarev:1986bi}.  The predicted number of PBHs relies sensitively on the scalar tilt of the primordial power spectrum as this sets the initial amplitude of the perturbations and PBHs are expected to form when the relative density contrast approaches the non-linear regime $\delta \rho / \rho \sim {\mathcal{O}}(1)$.  A matter or moduli dominated phase differs in some important ways compared to the typical estimate of PBH formation in a thermal history.  In particular, during a matter dominated phase there is no pressure to prevent the collapse of sub-horizon perturbations to evolving into black holes. This means that unlike the thermal case, perturbations can collapse to form PBHs long after crossing the Hubble radius -- the Jean's radius vanishes and so there is no pressure to prevent collapse. However, in \cite{Polnarev:1986bi} (see also \cite{doroshkevich1970spatial}) it was stressed that in a matter (dust) phase the resulting number of PBHs will depend on the probability for perfectly spherical collapse (so that angular momentum doesn't disrupt the process) and the duration of the period. In \cite{gsw} this analysis was adapted to the case of moduli domination (similar analysis with different motivation appeared in \cite{Carr:1994ar,Green:1997pr}).  One important consideration is that scalar fields on small scales can cease to be homogeneous and this can introduce a gradient pressure that prevents collapse \cite{hu1998structure,Hu:2000ke}. However, it was shown in \cite{gsw} that for the range of moduli masses of interest that this effect is negligible. It was also shown that the most relevant PBH constraint is requiring that PBH production does not exceed the critical density.  Given, a priori, the scalar tilt of the primordial power spectrum the corresponding constraints on the reheat temperature (and so moduli mass) were found in \cite{gsw}.  There it was also shown that for a fixed duration of moduli domination and a specified reheat temperature (both which follow from the moduli mass), constraints on the primordial scalar tilt $n_s$ can be established.  For high scale inflationary reheating and $T_r \simeq T_{BBN}$ the constraint $n_s\lesssim 1.3$ was found.  However, given that at the time CMB modes excited the horizon ($\sim 55$ e-foldings before the end of inflation) the scalar tilt is constrained to be strictly red ($n_s <1$) one can question the importance of this constraint.  This implies that only in models where significant changes in $n_s$ result would one be able to place strong constraints on the reheat temperature\footnote{This result may change given an extremely long period of matter domination, but it was found in \cite{gsw} that even for high scale inflationary reheating that requiring radiation domination by the time of BBN forces the epoch to be too short in duration to place constraints on a red spectrum.}. Examples of such the blue spectra on small scales have appeared in the literature, for example in models of hybrid inflation (see \cite{Young:2014oea} for a recent discussion of other possibilities and references).  In summary, PBH constraints do not lead to strong constraints on non-thermal histories unless the scalar tilt becomes blue ($n_s<1$) on sub-CMB scales.

\subsection{CMB Constraints \label{cmb_constraints}}
CMB observations are not only sensitive to the physics present during inflation, but also that during the post-inflationary history.
Significant work has gone into establishing predictions from inflation for both signatures in the temperature power spectrum and those associated with non-gaussianity (see e.g. \cite{Baumann:2014nda}).  In fact, it has been suggested that non-gaussianity arising during inflation may be useful in probing models with a split-SUSY spectrum \cite{Craig:2014rta}.  In this section we instead focus on the implications of the post-inflationary epoch on CMB observations.

\subsubsection{Consequences for Inflationary Constraints from the CMB}
\begin{figure}[p!]
\begin{center}
\includegraphics[scale=1.2]{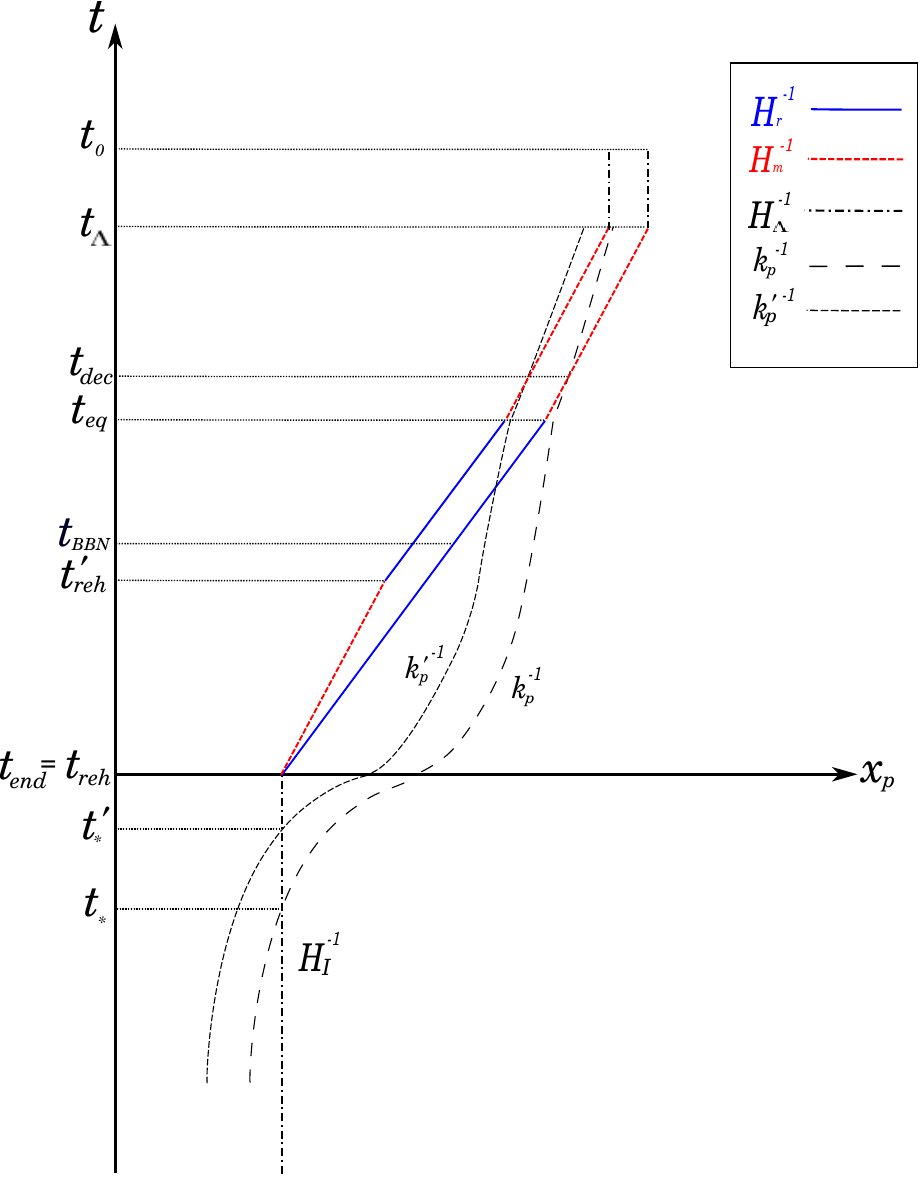}
\hspace{0.5cm}
\end{center}
\caption{Evolution of physical wavelengths as labelled by their inverse wavenumber $k_p^{-1}$ during inflation (below the x-axis) and during the post-inflationary epoch (above the x-axis). The solid (blue) line represents the Hubble radius, $\textcolor{blue}{H_r^{-1}}$ in a Universe dominated by a radiation fluid $w=1/3$, the dashed (red) line is the Hubble radius, $\textcolor{red}{H_m^{-1}}$ in a post-inflationary era dominated by a pressure-less fluid, $w=0$. We compare the evolution of a physical mode $k_*$  that re-enters at CMB decoupling in the standard scenario (Radiation $\rightarrow$ Matter $\rightarrow$ Dark energy) with a mode $k^{'}_{*}$ that re-enters at CMB decoupling in the nonthermal scenario (Matter$\rightarrow$ Radiation$\rightarrow$ Matter$\rightarrow$ Dark Energy).
These modes exit the Hubble radius at different times during inflation,  $t_*$ and $t^{'}_*$, which translates into a shift in the number of e-folds $\Delta N= H \Delta t$. The corresponding shift in the pivot scale or any co-moving mode is given by $k_*'=k_* e^{-\Delta N}$. \label{fig_rhb}}
\end{figure}
CMB data has reached an impressive level of accuracy and can be used to place stringent constraints on purposed inflation models.  However, 
to infer which models of inflation are most likely given the data, we match observations today to the predictions of the inflation model and this matching requires a knowledge of the post-inflationary expansion.  The equation of state (e.g. thermal $w=1/3$ or non-thermal $w=0$) during the post-inflationary epoch determines the expansion rate, and thus the rate at which primordial perturbations (re)enter the Hubble horizon.  For a spectral index $n_s$ that is not strictly scale-invariant ($n_s \neq 1$) this difference in expansion rate will lead to different predictions for the power spectrum \cite{Liddle:2003as,Kinney:2005in,Peiris:2008be,Adshead:2010mc,Mortonson:2010er,Easther:2011yq,Norena:2012rs,Martin:2010kz,Martin:2006rs,Cook:2015vqa}.  The matching and the differences that can arise due to two different post-inflationary histories are shown in Figure \ref{fig_rhb}.
As shown in the figure, given a mode of comoving wavenumber $k$, we must trace back to the instant this mode exited the Hubble horizon~\cite{Liddle:2003as,Kinney:2005in,Peiris:2008be,Adshead:2010mc,Mortonson:2010er,Easther:2011yq,Norena:2012rs,Martin:2010kz,Martin:2006rs}.  
This occurs when  $k=a_{k}H_{k}$ where $H$ and $a$ denote the Hubble parameter and scale factor respectively, and a subscript $k$ labels values at horizon crossing.  Where 
$\exp({N(k)}) \equiv {a_{end}}/{a_{k}}$ with $N$ the number of e-folds before the end of inflation and $a_{end}$ the scale factor at the end of inflation.
Assuming single-field slow-roll inflation for the purposes of illustration and characterizing the post-inflationary expansion by an effective equation of state $w$.  One finds the matching equation~\cite{Liddle:2003as}
\bea\label{matching}
  N(k,w) &\simeq& 71.21
        - \ln \left( \frac{k}{a_0H_0} \right )
        +\frac{1}{4} \ln \left( \frac{V_k}{m_p^4} \right) +\frac{1}{4} \ln\left( \frac{V_k}{\rho_{end}} \right) 
        + \frac{1-3w}{12\left( 1+w\right)} \ln\left( \frac{\rho_{{r}}}{\rho_{{end}}} \right),
\eea
where $\rho_{end}$ is the value of the energy density at the end of inflation, $V_k$ is the inflaton potential as the $k$th mode leaves the horizon, and $\rho_r$ is the energy density at which the universe is assumed to become  thermalized.  The first two terms in (\ref{matching}) are model independent. For GUT scale  inflation the third term is roughly $-10$. The   fourth term is typically order unity given that the value of the inflaton potential necessarily evolves slowly as inflation proceeds.   Finally, if the universe thermalizes promptly  the last term is negligible, and we recover the familiar result that  $50 \lesssim N \lesssim 60$ for modes contributing to the CMB.

In \cite{Easther:2013nga} the difference between a SUSY non-thermal history and thermal history was explored and using \eqref{matching} it was shown that the change in equation of state from the thermal case ($w=1/3$) to the moduli phase leads to a shift 
\be \label{deltaN}
 \Delta N = \fr{1-3w}{12(1+w)}\ln \left( \fr{ \rho_{r} }{ \rho_{end}} \right ), 
\ee
with  $\rho_{\rm{end}} = 3 V_{\rm{end}}/2$ for a given potential. If $w<1/3$ (in fact for the moduli phase we typically have $w=0$),  $\Delta N$ is negative, since  $\rho_{r} < \rho_{end}$ with $\rho_r$ the energy density at the end of the moduli phase.    
For simplicity we can assume inflationary (p)reheating was instantaneous and following \cite{Easther:2013nga}
we make the substitution $\rho_{end}\rightarrow\rho_{osc}^\sigma$, using $\rho_{r}^\sigma=(\pi^2/30) g_*(T_r^\sigma)(T_r^{\sigma})^4$. At the onset of oscillations $\rho_{osc}^{\sgm}(t_{osc})= \fr{1}{2}m^2_{\sgm}\Delta\sgm^2$ and \eqref{deltaN} becomes
\bea
\Delta N &=& -0.04 + \frac{1}{12}\ln \left( \frac{g_*(T_r^\sigma) T_r^4}{m_\sigma^2 \Delta\sigma^2} \right), \nonumber \\
&=&-10.75 + \frac{1}{12}\ln \left[       \left(\frac{g_*(T_r^\sigma)}{10.75} \right) \left( \frac{T_r}{3 ~\mbox{MeV}} \right)^4 
 \left(\frac{100 ~\mbox{TeV}}{m_\sigma} \right)^2 \left( \frac{m_p}{\Delta \sigma} \right)^2 \right], \label{thedeltaN} 
\eea
where we used $w=0$, and the second line expresses the parameters relative to fiducial values.  If  scalar decay proceeds via a gravitational strength coupling, equation~(\ref{decayrate}) eliminates the mass dependence in (\ref{thedeltaN}). With $c_3=1/(4\pi)$ we find
\be \label{deltaNred}
\Delta N=-10.68 + \frac{1}{18}\ln \left[       \left(\frac{g_*(T_r^\sigma)}{10.75} \right) \left( \frac{T_r}{3 ~\mbox{MeV}} \right)^4 
  \left( \frac{m_p}{\Delta \sigma} \right)^3 \right].
\ee 

This shift in the number of e-foldings leads to a shift in both the scalar tilt and tensor-to-scalar ratio $n_s$ and $r$. The uncertainty in $n_s$ is associated with the running $\alpha_s= dn_s/d\ln k$ and to lowest order in slow roll  \cite{Hoffman:2000ue,Schwarz:2001vv,Kinney:2002qn}
\bea\label{deltanr}
\nonumber \Delta n_s & = & \left. (n_s-1)\left[-\fr{5}{16}r-\fr{3}{64}\fr{r^2}{n_s-1}\right]\right| \Delta N,\\
\Delta r& = & \left. r \left[(n_s-1)+\fr{r}{8}\right] \right| \Delta N.
\eea 
The resulting fractional  uncertainties  $\Delta r/r, ~\Delta n_s/|n_s-1|$ in these  observables can be substantial \cite{Kinney:2005in,Adshead:2010mc}. In particular, the theoretical uncertainty in $n_s$ can be comparable to the precision with which it is measured by Planck \citep{Ade:2013rta}.  

The authors of \cite{Easther:2013nga} argued that accounting for additional consistencies of models, such as requiring a realistic dark matter candidate, could be used to improve the level of constraint.  As an example, it was shown that working in a SUSY framework, and appropriately accounting for the requirement of a successful WIMP candidate could lead to an improved situation for restricting inflation models.  As we discussed in Section \ref{dm_section}, late decays of the moduli (corresponding to low reheat temperatures which maximize $\Delta N$) will typically dilute thermal WIMPs to an unacceptable level for models with a long non-thermal epoch.  To remedy this it is usually assumed that WIMPs will be produced non-thermally in the decays.  However, this implies a large interaction cross-section and large predicted signals for indirect detection experiments in cases where all the dark matter is produced in this way.  In the MSSM this implies Wino or Wino/Higgsino like WIMPs and constraints from FERMI, PAMELA, and in some cases direct detection (such as Xenon100 and LUX) can place strong constraints on the cross-section.  The authors of \cite{Easther:2013nga} used these constraints to restrict the allowed reheat temperature, which in-turn reduces the uncertainties in $\Delta N$ and so the observables in \eqref{deltanr}.  The bounds on the reheat temperature were later improved in \cite{Cohen:2013ama,Fan:2013faa} with the inclusion of data from HESS and taking into account important effects such as Sommerfeld enhancement. We will discuss more about non-thermal dark matter and the constraints below.  Here we want to stress that although allowing for freedom in the post-inflationary history can lead to less stringent constraints on inflation model building when {\it only} CMB data is taken into consideration, properly accounting for other components (and the associated data sets) -- in this example we considered the origin and properties of dark matter -- leads to a new way to begin to probe the so-called dark ages prior to BBN.

\subsubsection{Dark Radiation and $N_{eff}$}
In addition to moduli decays to Standard Model particles and their superpartners, there may also be decays to hidden sector fields. 
Indeed, this is a common expectation in string-based models that give rise to the non-thermal history for dark matter discussed above
\cite{Higaki:2012ar,Cicoli:2012aq,Higaki:2012ba}. If the particles resulting from decay are light (meaning relativistic)
at the time of BBN and/or recombination, and non-interacting with MSSM particles, this leads to additional radiation coming from the hidden sector.
If these particles contribute substantially to the energy density they will affect 
the expansion rate changing predictions for both the abundances of primordial elements \cite{Steigman:2012ve} and the physics of the CMB~\cite{Abazajian:2013oma}.
Thus, using precision cosmological measurements one can establish constraints 
on the amount of dark radiation that is permitted within a particular class of models -- see \cite{Conlon:2013isa} and references within.

During radiation domination (after the decay of the last modulus) the effect of the hidden sector radiation on the Hubble expansion can be understood through the Hubble equation
$3 H^2 M_{pl}^2 = \rho_r$, where the total relativistic contribution
to the energy density is 
\be \label{basicr} 
\rho_r = \frac{\pi^2}{30}  g_\ast T^4,
\ee
with
\be \label{gdef}
g_\ast = g_{\mbox{\tiny MSSM}} + \sum_{i=bosons} g_i^h \left( \frac{T^h_i}{T} \right)^4 +\frac{7}{8} \sum_{i=fermions} g^h_i \left( \frac{T^h_i}{T} \right)^4
\ee
where the sums are over relativistic hidden sector particles with $g^h_i$ degrees of freedom, the factor of $7/8$ results from Fermi-Dirac statistics of fermions, $T^h$
is the temperature of the hidden sector particles (which importantly need not be equilibrated with Standard Model radiation), and $g_{\mbox{\tiny MSSM}} $ 
is the visible sector relativistic degrees of freedom, which in the early universe would be at least $g_{\mbox{\tiny MSSM}}=228.75$ in the MSSM, but near the MeV scale only  
the photons and neutrinos contribute with  
\be
g_{\mbox{\tiny MSSM}}(T) = g_\gamma + \frac{7}{8} g_\nu N_{\rm eff} \left( \frac{T_\nu}{T} \right)^4,
\ee
where $g_\gamma =2 $ for the photon, $g_\nu = 2$ for neutrinos, and $N_{\rm eff}$ is the effective number of neutrino species at temperature $T_\nu$.

At the time of BBN, the neutrino temperature tracks the photons so that $T_\nu = T$ and so with three relativistic neutrinos, $N_{\rm eff} =3$, the Standard Model prediction is $g_{\mbox{\tiny MSSM}}=7.25$.  However, because the neutrinos are weakly interacting $\langle \sigma_\nu v \rangle \sim G_F^2 T^2$
with $G_F \sim 10^{-5}$ GeV$^{-2}$ Fermi's constant, they decouple below the temperature of BBN ($\sim$ MeV) and the entropy in photons increases (so that the total entropy is conserved).
At the time of recombination we have
\begin{eqnarray}
g_{\mbox{\tiny MSSM}}(T_{rec}) &=& g_\gamma + \frac{7}{8} g_\nu N_{\rm eff} \left. \left( \frac{T_\nu}{T} \right)^4 \right\vert_{T=T_{rec}} \nonumber \\
  &=& 2 + \frac{7}{8} \cdot 2 \cdot \left(3.046 \right) \left( \frac{4}{11} \right)^{4/3} \nonumber \\
  &=&3.385,
\end{eqnarray}
where $N_{\rm eff}=3.046$ for the Standard Model \cite{Dolgov:2002wy}. 
The increase in the photon entropy following neutrino decoupling leads to an increase in the temperature of $T_\nu / T=(4/11)^{1/3}$ \cite{Abazajian:2013oma}. Given the Standard Model (or MSSM) predictions, any new dark radiation would lead to an additional contribution to $g_\ast$ or 
$N_{\rm eff}$.  For historic reasons, constraints on new hidden sector radiation are typically expressed through $N_{\rm eff}$.
In applying these constraints its important to remember that  
if hidden sector radiation couples differently to decaying moduli (or the inflaton during reheating) than Standard Model particles, this 
can lead to different temperatures for each species and this will be preserved in the absence of interactions between the systems of particles.
As we will discuss soon, such a situation can also lead to isocurvature perturbations. The connection between isocurvature and dark radiation in non-thermal histories was studied in \cite{Iliesiu:2013rqa}.

Constraints from BBN result from requiring agreement with both the abundances of $^4He$ and $D$ \cite{Cyburt:2004yc}, which implies $N_{\rm eff}=3.24 \pm 1.2$ at the time of BBN.
At the time of recombination, the Planck satellite provides constraints with the current results  \cite{Planck:2015xua}  implying 
$N_{\rm eff}=3.13 \pm 0.32$ ($68\%$ CL), which is consistent with $\Delta N_{eff}=0$.   
\\

{\bf Dark Radiation and UV Completions.}
We have seen that axions are a generic prediction of string / M-theory and as such light axions can lead to strong constraints in light of bounds on $N_{eff}$.
The consequences have been explored in both the case of M-theory \cite{Acharya:2010zx}, Type IIB Large Volume string compactifications \cite{Queiroz:2014ara,Cicoli:2013ana,Conlon:2013isa,Cicoli:2012aq,Hebecker:2014gka}, and more generally \cite{,Iliesiu:2013rqa,Honecker:2013mya,Fairbairn:2013gsa,Higaki:2013lra,Higaki:2012ar}.
If these axions remain light (i.e., if the shift symmetry holds to a good approximation after moduli stabilization), then they could constitute dark radiation. Two questions thus naturally arise: $(i)$ how generic is the presence of dark radiation in string compactifications?, and $(ii)$ if both dark radiation and dark matter have a common origin from the decay of a string modulus, can one find a correlation between them by combining dark matter observational constraints with present bounds on $N_{\rm eff}$? These questions have been addressed in \cite{Iliesiu:2013rqa,Allahverdi:2014ppa, Cicoli:2012aq}.

For question $(i)$, we note that in string compactifications with perturbative moduli stabilisation (like the Large Volume Scenarios), the production of dark radiation is unavoidable. This is easy to see, since the axionic shift symmetry is protected at perturbative level and broken only by non-perturbative effects. Thus, while a modulus gets a large mass during perturbative stabilization, the axionic directions are only slightly lifted by non-perturbative effects\footnote{Light axions can still be removed from the low-energy spectrum by being `eaten' by the anomalous $U(1)$ gauge bosons due to the anomaly cancellation. However, this is not the case for light closed string axions living in the bulk. In stabilization schemes that dominantly depend on non-perutrbative contributions to the superpotential (like KKLT), the axionic direction is strongly lifted.}.

With regards to question $(ii)$, in the presence of dark radiation, the reheating temperature $T_{r}$ can be written
as a function of $N_{\rm eff}$ and the mass of the modulus $m_{\sigma}$.  
That is, the modulus decays to both visible and hidden degrees of freedom with partial decay widths
given by $\Gamma_{\rm vis} = c_{\rm vis} \Gamma_0$ and $\Gamma_{\rm hid} = c_{\rm hid} \Gamma_0$, respectively, 
where $c_{\rm vis}$ and $c_{\rm hid}$ are model-dependent prefactors and, from (\ref{decayrate}), we see that $\Gamma_0 = (1/2\pi)(m^3_{\sigma}/m^2_p)$. Since the hidden sector is composed of light axions with only gravitational couplings, these relativistic decay products
do not thermalise and in fact add to the effective neutrino species.

The expression for $\Delta N_{eff}$ may be obtained as follows. The conservation of comoving entropy $s = g(T) a^3 T^3$ implies that the temperature and energy density of the visible sector go as $T_{vis} \sim \frac{1}{g^{1/3} a}$ and $\rho_{vis} = \frac{\rho_{vis}^{init}}{g^{1/3} a^4}$. Then, at neutrino decoupling, the ratio of the energy densities is given by 
\be
\frac{\rho_{hidden}}{\rho_{vis}} = \frac{\rho_{hidden}^{init}}{\rho_{vis}^{init}} \left( \frac{g(T_{dec})}{g(T_{reheat})} \right)^{1/3}.
\ee
The number of extra effective neutrino species is then
\bea \label{delneffepr}
\Delta N_{eff} & = & 3 \frac{\rho_{hidden}}{\rho_{neutrinos}} =  \frac{43}{7} \frac{\rho_{hidden}}{\rho_{vis}} \nn \\
&  = & \frac{43}{7} \frac{c_{\rm hid}}{c_{\rm vis}} \left( \frac{g(T_{dec})}{g(T_{reheat})} \right)^{1/3}.
\eea

On the other hand, the visible sector reheats to a temperature given by (\ref{Tr}), with $c$ replaced by $c_{\rm vis}$. Using (\ref{delneffepr}) to eliminate $c_{\rm vis}$ of (\ref{Tr}) in favor of $\Delta N_{eff}$ and $c_{\rm hid}$, one obtains the parametric relation between $T_r$ and $\Delta N_{eff}$:
\be \label{trneffrel}
T_r \, \sim \, \frac{1}{\sqrt{\Delta N_{eff}}} \,\,.
\ee
The constant of proportionality depends on $c_{\rm hid}$ and $m_{\sigma}$ (for the full expression, we refer to \cite{Allahverdi:2014ppa}). For $c_{\rm hid}=1$, $g_*=68.5$ and $m_\sigma = 5\cdot 10^6$ GeV, requiring $\Delta N_{\rm eff} \lsim 0.96$ translates into a lower bound on the reheating temperature: $T_{\rm rh} \gsim 0.73$ GeV. 

There is one striking consequence of (\ref{trneffrel}). As $\Delta N_{eff} \rightarrow 0$, the reheat temperature becomes large enough that any dark matter has time to thermalize after being produced. Thus, \textit{if future observations reveal that the effective number of neutrino species is close to the Standard Model value, the implication for dark matter is that a thermal history seems preferred.} This is non-trivial information about the thermal history of dark matter coming from the CMB.

\begin{figure}[t!]
\begin{center}
\includegraphics{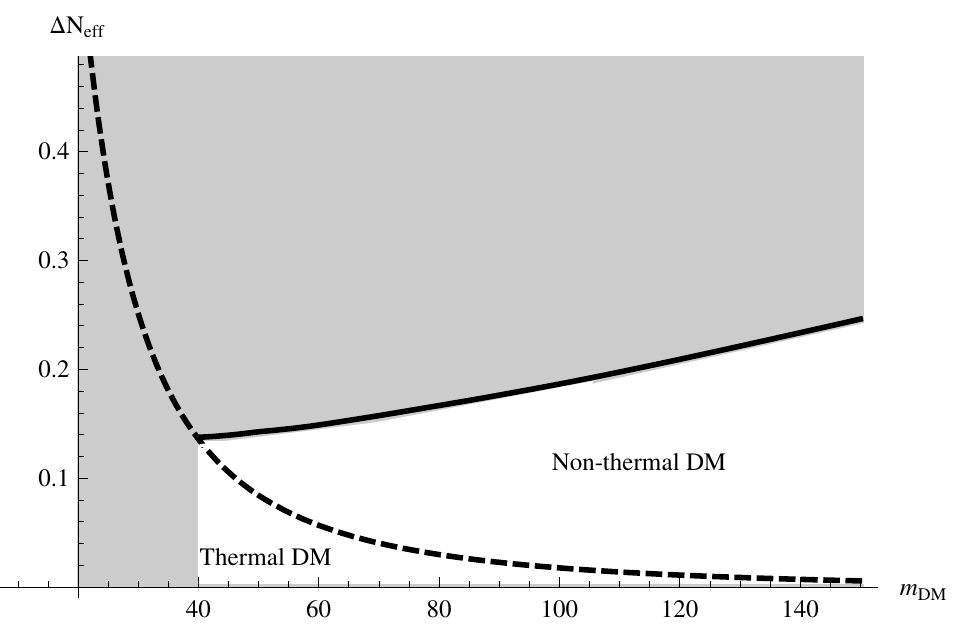}
\hspace{0.4cm}
\end{center}
\caption{Constraints on the $(\Delta N_{\rm eff},m_{\rm DM})$-plane for $c_{\rm hid}=1$,
$g_*=68.5$ and $m_\phi= 5\cdot 10^6$ GeV: the solid line is based on Fermi data
whereas the dashed line represents the freeze-out temperature. The shaded region is ruled out due to DM overproduction
both in the thermal case (for $m_{\rm DM}\lesssim 40$ GeV and below the dashed line) and in the non-thermal branching scenario
(above the solid and dashed lines).\label{figdrneff1}}
\end{figure}

\begin{figure}[t!]
\begin{center}
\includegraphics{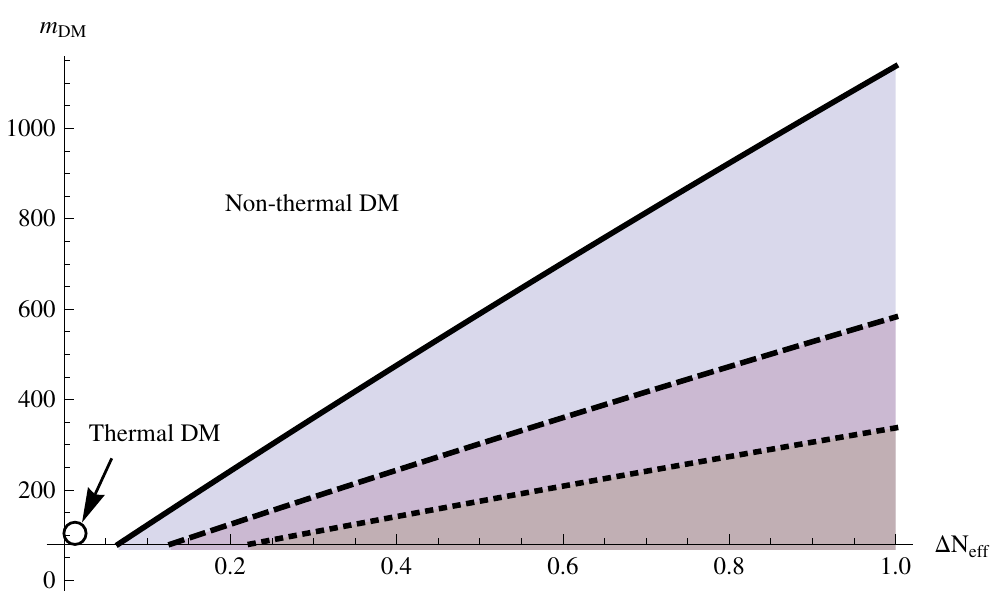}
\hspace{0.4cm}
\end{center}
\caption{Lower bound on the DM mass as a function of $\Delta N_{\rm eff}$ for different values of the modulus mass:
$m_\phi=4\cdot 10^6$ GeV (solid line), $m_\phi= 5\cdot 10^6$ GeV (dashed line) and $m_\phi= 6\cdot 10^6$ GeV (dotted line).
The shaded region is ruled out based on Fermi data.
Here we have set $g_*=68.5$ and $c_{\rm hid}=1$.\label{figdrneff2}}
\end{figure}

One can combine (\ref{trneffrel}) with bounds on the annihilation cross section of dark matter coming from Fermi data (we will use the dwarf galaxy constraints described in Section \ref{dm_section}). To do this, it is useful to re-express the Fermi constraints in terms of the reheat temperature. This can be done as follows. Gamma-ray data places an upper bound on the annihilation cross section for given final states as a function of the dark matter mass. Assuming that the relic density is satisfied in a non-thermal scenario and is given by (\ref{TandNTrelation}), and using the fact a typical freeze-out temperature is $T_f \sim m_{\chi^0}/20$ GeV, one can re-express the constraints as a \textit{lower} bound on $T_r$ as a function of the dark matter mass. Using (\ref{trneffrel}), this then translates into an \textit{upper} bound on $\Delta N_{eff}$ as a function of dark matter mass. This can be seen in Figure \ref{figdrneff1}.
In Figure \ref{figdrneff2}, these same exclusion plots have been displayed for various values of the modulus mass.

\subsubsection{Isocurvature and the CMB}
Consistency with cosmological observations requires that the largest contribution to the temperature anisotropy of the CMB must be adiabatic \cite{Planck:2015xua}.
Rigid constraints can be placed on models that predict departures from strict adiabaticity, however the level of constraint can be model dependent as multiple isocurvature components and theoretical priors can play an important role \cite{Bucher:2000hy}. For a single source of isocurvature Planck places strong constraints on model building where isocurvature can be ruled out at the percent level \cite{Planck:2015xua}.  

In some cases non-thermal histories with moduli can lead to a departure from adiabaticity and significant isocurvature contributions.  If the fields during inflation have masses comparable to (or less than) the Hubble scale, the moduli can fluctuate and lead to a second source of cosmological perturbations -- differing from the inflaton. 
This spectrum would be different than that of the inflaton (and its decay products) and so depending on the post-inflationary history this could lead to a large source of isocurvature perturbations at the time of recombination. For the histories with moduli we have considered in this review, an analysis of isocurvature constraints and their importance for non-thermal histories was performed in \cite{Iliesiu:2013rqa} (for earlier studies see \cite{Lemoine:2009is,Lemoine:2009yu}). 
To understand the possible origin of isocurvature in non-thermal histories consider the curvature perturbation associated with each type of fluid in the early universe (e.g. radiation, dark matter, moduli, etc.).
The curvature perturbation of the $i$th fluid is
\be
\zeta_i = -\Psi-H\frac{\delta\rho_i}{\dot{\rho}_i} \, .
\ee
where $\Psi$ is the Newtonian potential and dots denote time derivatives, and $\delta \rho_i$ is the density fluctuation of the $i$th fluid. 
In connecting with observations it is conventional to define the isocurvature contribution of a particular fluid relative to 
the radiation perturbation so that we define the (gauge invariant) isocurvature perturbation of the $i$th fluid by (e.g. \cite{Malik:2008im})
\be \label{entropyr} 
S_{i} = 3 \left( \zeta_i - \zeta_r \right) \, ,
\ee
where $\zeta_r$ is the radiation density perturbation.  Observables are given by the correlation functions of these fluctuations or more conventionally by the power spectrum
defined as
\be
\langle{X(\mathbf k) Y(\mathbf k')}\rangle = (2\pi)^3 \delta^3(\mathbf k - \mathbf k')P_{XY} \, , 
\ee
for two fluctuations $X$ and $Y$.  The observables for the isocurvature fraction, $\alpha_i$, its correlation with the curvature perturbation, $r_i$, and the cross correlation between any two isocurvature modes, $r_{ij}$ are given by 
\begin{align} \label{stuff}
\alpha_i&= \frac{P_{S_iS_i}}{\sum_jP_{S_jS_j}+P_{\zeta\zeta}} \, , \\
r_i&=\frac{P_{S_i\zeta}}{\sqrt{P_{S_iS_i}P_{\zeta\zeta}}} \, , \\
r_{ij}&=\frac{P_{S_iS_j}}{\sqrt{P_{S_iS_i}P_{S_jS_j}}} \, , \label{stuff2}
\end{align}
which are all quantities that are constrained by observations \cite{Planck:2015xua}.

{\it When do isocurvature perturbations arise in non-thermal histories?}
If the mass of the modulus is lighter than the Hubble scale $m_\sigma  \lesssim H_I$ during inflation,
then the quasi-deSitter period will result in long-wavelength fluctuations of the modulus field with an average amplitude 
$ \delta \sigma \sim {H_I}/{2\pi}$.
This leads to an additional source of cosmological perturbations with the corresponding curvature perturbation different from that of the inflaton. 
Following inflationary reheating -- where in a thermal history all of the energy and matter of the universe is assumed to be created -- the modulus can decay leading to an additional source of radiation and matter. Thus, whereas radiation and matter created during inflationary reheating will inherit the inflaton's fluctuation $\zeta_I$, those produced from moduli decay will instead be set by the modulus ($\delta\sigma$) and so \eqref{stuff}-\eqref{stuff2} may be non-zero leading to constraints. 

However, even though we have seen that moduli decay is an essential part of a non-thermal history, there are still many ways in which isocurvature perturbations may be observationally irrelevant and lead to no new constraints on model building. Firstly, if the mass of the modulus is above the Hubble scale during inflation $m_\sigma  > H_I$, which may be argued to be generic \cite{Baumann:2011nk,Avgoustidis:2012yc}, then in a single Hubble time the amplitude of its fluctuations will be exponentially suppressed on large scales by a factor $\exp(-m_\sigma^2 / (3 H_I^2) )$ and so the inflaton will be the only relevant source of cosmological fluctuations \cite{Linde:2005ht}. Another important observation was made by Weinberg, who demonstrated that even if an isocurvature mode is generated initially, if local thermal equilibrium is reached these modes will become adiabatic  \cite{Weinberg:2004kf}. And finally, if the modulus comes to dominate the energy density of the universe (determining the cosmic expansion rate) this also has the effect of washing out any existing isocurvature perturbations.  In most cases studied thus far arising in string and M-theory negligible isocurvature results \cite{Iliesiu:2013rqa}.

\section{Challenges and Outlook}
We have seen that the presence of moduli in the early universe, along with SUSY breaking near the TeV scale, motivate a non-thermal history.
We discussed how this approach can account for cosmological dark matter, axions as a solution to the strong CP problem, and the baryon asymmetry of the universe.
However, we have also seen that in many cases existing data already puts very tight constraints on each of these.
The most severe case is for standard SUSY WIMPs \cite{Fan:2013faa, Cohen:2013ama} suggesting that more creative dark matter model building in the hidden sector may be necessary (see e.g. \cite{Blinov:2014nla}).
We have also seen that a number of cosmological constraints -- particularly those associated with the CMB and structure formation -- can lead to rigid constraints on these scenarios.  
As data continues to improve at a rapid pace, this should help us continue to work toward unmasking the cosmological history of the post-inflationary universe and shape the direction for model building.

In most of this review we have emphasized the model independent aspects of non-thermal histories arising from moduli.
This is possible because the prediction relies on the physics of the hidden sector.  However, a particular challenge for string/M-theory approaches is embedding the visible sector.
This is a challenging endeavor, requiring a model dependent approach, but we emphasize here that in cases where this is possible and quasi-realistic models can be constructed that the physics is largely independent of the hidden sector -- a non-thermal history remains.
Other model dependent challenges do arise.  The `gravitino problem' is a common example.
Gravitinos produced from the decaying modulus can lead to both dark matter and dark radiation production.  This will depend on model dependent features, such as the branching ratio and the gravitino mass.  Whether this is a problem or not, depends on the model.  For example, in the $G2-MSSM$, the gravitino mass is comparable to the moduli mass and so the decay is kinetically forbidden \cite{Acharya:2008bk}.  Another challenging problem is the so-called `over-shoot' or `Kallosh-Linde' problem \cite{Kallosh:2004yh}.  As an example, in Type IIB (KKLT) string compactifications, the height of the stabilization barrier is set by the gravitino mass.  Thus, if the Hubble scale during inflation exceeds this barrier height the moduli can overshoot their minimum and de-compactify the extra dimensions.  This problem is again model dependent and depends on the stabilization mechanism.  For example, in models of `Strong Moduli Stabilization' the problem can be avoided \cite{Dudas:2012wi,Garcia:2013bha,Evans:2013nka}.  A more model independent solution may result by having a dynamically changing gravitino mass as discussed in \cite{He:2010uk}.
Model dependent obstructions like the overshoot and gravitino problems emphasize the need for further theoretical investigation.  Regardless, we have seen that a non-thermal history is a robust prediction of hidden sector physics in string / M-theory approaches to beyond the Standard Model physics.

\section*{Acknowledgements}
This research was supported in part by NASA Astrophysics Theory Grant NNH12ZDA001N, and DOE grant DE-FG02-85ER40237. 
S.W. would like to thank McGill University, the University of North Carolina, Northeastern University, the University of Utah, and the Michigan Center for Theoretical Physics for hospitality while this work was completed.

\bibliographystyle{utphys}

\end{document}